\documentclass[preprint,12pt]{elsarticle}
\usepackage[a4paper, left=2cm, right=2cm]{geometry}
\usepackage{amssymb}
\usepackage{amsmath}
\usepackage{comment}
\usepackage[normalem]{ulem}  % for strike-through support

\let\oldAA\AA
\renewcommand{\AA}{\text{\oldAA}}
\newcommand{\uu}[1]{\ensuremath{\,{\mathrm{#1}}}}
\usepackage{xspace}
\usepackage{multicol}
\usepackage{xcolor}
\usepackage[flushleft]{threeparttable}
\usepackage{gensymb}
\usepackage{textcomp,mathcomp}
\usepackage{url}
\newcommand{\etal}{\emph{et al.}\xspace}

\newcommand{\add}[1]{#1}
\newcommand{\del}[1]{#1}

\journal{Acta Materialia}

\begin{document}

\begin{frontmatter}

\title{Interplay between alloying and tramp element effects on temper embrittlement in bcc iron: DFT and thermodynamic insights}

\author[inst1]{Amin Sakic}

\affiliation[inst1]{organization={Department of Materials Science, Montanuniversität Leoben},
            addressline={Franz Josef-Straße 18}, 
            city={8700 Leoben}, 
            country={Austria}}
\affiliation[inst2]{organization={Christian Doppler Laboratory for Knowledge-based Design of Advanced Steels, Department of Materials Science, Montanuniversität Leoben},
            addressline={Franz Josef-Straße 18}, 
            city={8700 Leoben}, 
            country={Austria}}
\author[inst2]{Ronald Schnitzer}
\author[inst2]{David Holec}

\begin{abstract}
The details of the temper embrittlement mechanism in steels caused by impurities are unknown. Especially from an atomistic point of view, there are still open questions regarding their interactions with alloying elements such as Ni, Cr, and Mo. Therefore, we used density functional theory to investigate the segregation and co-segregation behavior and the resulting influence on the cohesion of three representative tilt grain boundaries in iron. The results are implemented in a multi-site and multi-component kinetic and thermodynamic model for grain boundary segregation, to gain insights into the temporal and final grain boundary coverage. Our results show that the segregation tendency of As, Sb, and Sn is stronger than that of the alloying elements and significantly mitigates the grain boundary cohesion. Depending on the GB type, interactions between Sb and Sn vary from negligible to strongly attractive, which increases the likelihood of co-segregation. The cohesion-weakening effect is further amplified when elements such as Sb, Sn, and As co-segregate, compared to their individual segregation. In contrast, the co-segregation of Ni and Cr does not significantly increase the enrichment of impurities at grain boundaries, and their impact on cohesion is found to be negligible. The ability of Mo to mitigate reversible temper embrittlement is primarily attributed to its cohesion-enhancing effect and its capability to repel tramp elements from GBs, rather than scavenging them within the bulk, as suggested by previous literature. 
\end{abstract}

% \includegraphics[width=\textwidth]{graphical_abstract_final.pdf}

%% Add research highlights
% \begin{highlights}
% \item \add{Interplay between tramp elements (Sn, Sb, and As) and alloying elements (Cr, Mo, and Ni) on the mechanism of reversible temper embrittlement was investigated with DFT in conjungtion with thermodynamic and kinetic simulations.}
% \item \add{Sn, Sb, and As show strongest segregation tendencies and detoriate GB cohesion substantially.}
% \item \add{Repulsive bulk interactions indicate that Mo does not prevent RTE through a scavenging effect, but rather due to the ability of Mo to increase GB cohesion and dependent on GB type shield the GB from Sn, Sb, and As through repulsive interactions.}
% \end{highlights}

% \begin{keyword}
% %% keywords here, in the form: keyword \sep keyword
% DFT \sep temper embrittlement \sep tramp elements \sep kinetic and thermodynamic simulations%% PACS codes here, in the form: \PACS code \sep code
% %\PACS 0000 \sep 1111
% %% MSC codes here, in the form: \MSC code \sep code
% %% or \MSC[2008] code \sep code (2000 is the default)
% %\MSC 0000 \sep 1111
% \end{keyword}

\end{frontmatter}


\section{Introduction}
\label{sec:sample1}
Originally discovered in 1883 and extensively investigated through experimental methods, the underlying mechanism of reversible tempering embrittlement (RTE) is enigmatic. This phenomenon, commonly observed in low alloy steels featuring martensitic, bainitic, pearlitic, or ferritic microstructures, leads to diminished ductility, toughness, and consequently fracture resistance. Following tempering or gradual cooling within a critical temperature range of 350-650$\,$°C \cite{Olefjord1978-df}, RTE transforms the low-temperature fracture mode from cleavage to intergranular. Even before the accumulation of experimental data, the pioneering theoretical insights of McLean \cite{mclean1958grain} hinted at the potential association between intergranular embrittlement and the segregation of solute atoms along grain boundaries (GBs). In 1956, Balajiva and collaborators \cite{balajiva1956effects} identified a correlation between RTE and specific impurity elements (P, As, Sb, Sn). The prevailing consensus emerged, attributing this behavior to the segregation of these elements at GBs, thereby diminishing cohesion. Subsequent research by Low \etal \cite{low1968alloy} unveiled the pivotal role of alloying elements in the RTE phenomenon. Their study demonstrated that plain C steels doped with impurities do not exhibit susceptibility to RTE, while Ni-Cr steels do. In contrast, their findings also highlighted that the inclusion of Mo effectively reduces the susceptibility of Ni-Cr steels to RTE.
By far the most extensively studied impurity in the past has been P~\cite{mulford1976temper}, owing to its prevalence in significantly higher quantities in most commercial steels compared to Sb, Sn, and As. Consequently, P was commonly considered the primary factor behind RTE.
To elucidate the co-segregation phenomena of alloying and impurity elements, Guttmann introduced a thermodynamic formalism based on the regular solution model, incorporating interactions among the segregating species~\cite{guttmann1975equilibrium}. He posited that strong, attractive bulk interactions between P and Mo would give rise to a scavenging effect, consequently diminishing the segregation of P to GBs. Furthermore, he suggested that attractive GB interactions between alloying elements and impurities promote the GB enrichment of these constituents. Similar conclusions were reached in various studies exploring the influence of Sb and Sn in Ni-Cr steels~\cite{ohtani1976temper, cianelli1977temper}. However, subsequent research has revealed that there are no discernible direct interactions between Mo and P, and the GB segregation of P remains unaffected by the presence of Ni, Cr, and Mn~\cite{briant1981effect, grabke1987effects}. The role of P in causing RTE has been the subject of frequent investigation through various theoretical approaches, particularly atomistic simulations~\add{\cite{mai2023phosphorus,wang2021first}}. For a comprehensive overview, we refer the reader to a recent ab initio informed publication by H. Mai~\cite{mai2023phosphorus}. In their work, they concluded that the presence of P as an individual alloying element does not significantly impair GB cohesion, and therefore, it cannot be considered the sole cause of P-induced RTE. Moreover, their findings revealed that only the combination of P and Mn has a more detrimental effect on GB cohesion compared to their individual impacts. From a theoretical perspective, the behavior of residual elements such as Sb, Sn, and As remains a relatively underexplored topic. For instance, based on experimental data on the enthalpy and entropy of GB segregation, Lejcek \etal~\cite{lejvcek2021entropy} concluded that impurity elements like Sn, Sb, and P exhibit a preference for interstitial over substitutional GB sites at finite temperatures. Only recently, Rehak \etal~\cite{vrehak2023role} demonstrated that the vibrational energy contribution does not alter the substitutional site preference of Sn and Sb in the $\Sigma 5 (310)[001]$ Fe GB obtained from density functional theory (DFT) at $0\,$K.
However, to gain a deeper understanding of the elemental effects, it is imperative to evaluate the segregation profiles of the investigated impurities and their impact on GB cohesion in a broader range of GB types. This is particularly important, as it has been demonstrated that GB type strongly influences the segregation and cohesion effects.
Additionally, the co-segregation with commonly used alloying elements like Cr, Ni, and Mo, and their combined impact on GB cohesion, continue to be unresolved issues. This becomes especially relevant in the context of the steel industry's transition from ore-based to scrap-based production methods, as the prevalence of these elements is expected to significantly increase~\cite{dworak2023stahlrecycling, daigo2021potential}. \add{We note that in this work, we do not include C, as it is well-documented in the literature~\cite{wang2021first, seah1975interface, seah1976segregation} that small solutes like C and B have a strong segregation tendency to GBs, by that keep the embrittling elements away from the GB,  and significantly contribute to GB strengthening. Therefore, they are potent solutes for mitigating RTE. As such, our simulations are valid for interstitial-free (IF) steels and ultra-low carbon steels.} 

Therefore, we employed DFT to compute the segregation and co-segregation propensities, as well as the impact of Sn, Sb, As, Cr, Mo, and Ni on the cohesion of three distinct body-centered cubic (bcc) Fe GBs. Our study began by determining the segregation profiles and their impact on the cohesion of each solute within each GB type. Subsequently, we assessed the co-segregation tendencies and solute interactions within each GB. Notably, we systematically altered the position of the second segregation species, keeping the position of the first solute fixed at its minimum single segregation site. The resulting configuration with minimal energy was then utilized to assess the influence of co-segregation on GB cohesion, which was then compared to the individual effects. In the final section, our attention shifted to the assessment of time-dependent GB enrichments and those achievable in the thermodynamic limit, utilizing analytical models built upon the atomistic inputs.

\section{Methodology}
\label{sec:methodology}
The quantum-mechanical calculations were conducted utilizing DFT~\cite{hohenberg1964density, PhysRev} within the framework of the Vienna Ab-initio Simulation Package (VASP)~\cite{Kresse1996-gt, Kresse1996-tg}. The projected augmented wave (PAW) method~\cite{Kresse1999-if} was employed for accurate electron-ion interactions. The exchange-correlation potential was described using the generalized gradient approximation (GGA) with the Perdew, Burke, and Ernzerhof (PBE) parameterization~\cite{Perdew1996-vd}. The pseudopotentials employed for electron-ion interactions treated both the valence electrons and the $p$ closed-shell electrons as valence electrons for Fe, Mo, Cr, and Ni elements. All calculations were carried out under spin-polarized conditions. A plane wave cutoff energy of $500\uu{eV}$ was chosen for the expansion of the wave functions. To sample the first Brillouin zone, an automatic $k$-mesh generation method was employed with a length parameter ($R_k$) set at $50\uu{\AA}$, resulting in a $\Gamma$-centered mesh. This approach yielded between 6000 and 14500 $k$-points$\cdot$atom within the first Brillouin zone. Integration over the $k$ points was performed using a first-order Methfessel-Paxton scheme with a thermal smearing parameter of $0.2\uu{eV}$. Convergence was ensured by setting the ionic relaxation criterion to below $10^{-4}\uu{eV}$. These parameter choices were guided by convergence tests, with the total energy changes converging within $\approx 1\uu{meV/at.}$ to maintain accuracy and consistency. \add{These settings lead to all force components being $\lessapprox 0.01\uu{eV/\AA}$.}

\subsection{\del{Grain boundary structures}\add{Structural models}}
We considered three GBs, namely the $\Sigma 3 (1\bar{1}1)[110]$, $\Sigma 3(1\bar{1}2)[110]$, and $\Sigma 9(2\bar{2}1)[110]$ symmetric tilt GBs, as depicted in Fig.~\ref{fig:seg_profiles}. These structures were extensively employed in previous studies regarding elemental segregation in bcc Fe, making them well-suited for comparative analyses. For simplicity, we will henceforth refer to them as $\Sigma 3 (1\bar{1}1)$, $\Sigma 3(1\bar{1}2)$, and $\Sigma 9(2\bar{2}1)$, respectively. From a structural standpoint, the $\Sigma 3 (1\bar{1}1)$ GB epitomizes a characteristic high-angle GB, while the $\Sigma 9(2\bar{2}1)$ GB exhibits attributes of a low-angle GB due to its atomic configuration resembling an edge-dislocation core at the GB plane~\cite{bhattacharya2014ab}. Analogous to stacking faults or twins, the $\Sigma 3(1\bar{1}2)$ GB shares similarities with the $\Sigma3 (111)$ GB found in fcc metals. It is worth emphasizing that symmetric tilt GBs with a rotation axis aligned along the $\langle110\rangle$ direction are more prevalent in polycrystalline Fe than other orientations~\cite{bhattacharya2014ab}. The construction of all GB structures was carried out using the \textit{build} module within the python Atomic Simulation Environment (ASE) package~\cite{larsen2017atomic}. To this end, we adopted the fully relaxed bcc (ferromagnetic) Fe bulk with a lattice parameter of $2.839\uu{\AA}$. To ensure stability during calculations and maintain consistency with prior literature \cite{mai2022segregation, larsen2017atomic}, atoms closer than $0.7$ times the bulk lattice constant were merged into a single atom, located at the mean position of the original pair. To avoid issues and attain convergence, a vacuum layer of at least $6\uu{\AA}$ was added atop each structure based on surface energy convergence tests. In our setup, the cell shape and volume were held constant, while atomic positions were subjected to relaxation. This arrangement allowed the relaxation of the GB distance to be absorbed by the vacuum. Specifically, the GB structures employed for calculating the work of separation ($W_{\text{sep}}$) were separated by $6\uu{\AA}$ at the designated plane of interest. \add{Bulk calculations were conducted using a $4\times4\times4$ bcc Fe supercell, containing 128 atoms, to circumvent the interactions between the solute atom and its periodic boundary image.}

\subsection{Grain boundary energy and excess volume}
The GB energy ($\gamma_{GB}$) is the excess energy when two adjacent crystals of the same material but different orientations are joined. We construct a slab with one GB of interest (in the middle) and two free surfaces. The corresponding GB energy, $\gamma_{GB}$ is calculated as
\begin{equation}
    \gamma_{GB} = \frac{E_{GB}-E_{FS}}{A}\ ,
\end{equation}
where $E_{GB}$ and $E_{FS}$ represent the total energies of the GB and free surface slabs (containing the same number of atoms and the same geometry of the surfaces), respectively, and $A$ is the cross-sectional area. We acknowledge that GBs can be modeled using various approaches. However, as demonstrated in Ref.~\cite{scheiber2016ab}, the setup with a vacuum converges $\gamma_{GB}$, $\gamma_{FS}$, and the work of separation ($W_{sep}$) most rapidly with respect to grain thickness. Since GBs exhibit varying degrees of disturbance in perfect stacking, depending on the GB type, the specific volume (per atom) typically increases in the GB regions compared to the bulk. This volume increase normalized by the GB area is described by the excess volume ($V_{exc}$), which hence has the length dimension. We calculate $V_{exc}$ by extracting it from the relaxed GB and surface structures as the difference between the outermost atoms, following the approach used in Ref.~\cite{scheiber2016ab}.

\subsection{Segregation}
\subsubsection{Segregation energy}
The segregation energy ($E_{seg}(X)$) quantifies the energy difference between a solute $X$ located in the bulk and at a GB site. It is calculated as follows: 
\begin{equation}
    E_{seg}(X) = E_{GB}[(n-1)\mathrm{Fe},X] + E_{bulk} - E_{GB} - E_{bulk}[(\del{n}\add{m}-1)\mathrm{Fe},X]
\end{equation}
where $E_{GB}[(n-1)\mathrm{Fe},X]$ and $E_{bulk}[(\del{n}\add{m}-1)\mathrm{Fe},X]$ are the total energies of simulation boxes with one solute at a GB and a bulk site, respectively. \add{The symbols $n$ and $m$ denote the number of atoms in the GB and bulk cell, respectively.} A negative $E_{seg}$ indicates that the segregation from bulk to GB is energetically preferable.

\subsubsection{Incremental segregation energy}
The incremental segregation energy $E_{seg}^{inc}(Y|X)$ gives the likelihood of a second solute, $Y$, segregating to a GB where solute atom $X$ has already occupied the energetically most preferable site at the GB. 

\begin{equation}
    E_{seg}^{inc}(Y|X)=E_{GB}\left[\left(n-2\right)\mathrm{Fe}, X_i, Y_j\right]-E_{GB}\left[\left(n-1\right)\mathrm{Fe}, X_i\right]-E_{bulk}\left[\left(m-1\right)\mathrm{Fe}, Y\right]+E_{bulk}
    \label{eq:E_seg^inc}
\end{equation}

Here, $E_{GB}\left[\left(n-2\right)\mathrm{Fe}, X_i, Y_j\right]$ is the energy of the GB slab with two atoms placed at distinct GB-sites: atom X is placed on the GB site $i$, which is the preferred substitutional site for that atom, followed by the atom Y being placed on the GB site $j$, such that the term $E_{GB}\left[\left(n-2\right)\mathrm{Fe}, X_i, Y_j\right]$ is minimized.

\subsubsection{Interaction energy}
The interaction energy\add{, often referred to as binding energy,} provides information on the type and magnitude of the interactions between solute $X$ and solute $Y$. We further distinguish between both solutes occupying GB sites (Eq.~\eqref{eq:int_GB} or being in the bulk (Eq.~\eqref{eq:int_bulk}):
\begin{gather}
    E_{int}^{GB} = E_{GB}\left[\left(n-2\right)\mathrm{Fe}, X_i, Y_j\right] - E_{GB}\left[\left(n-1\right)\mathrm{Fe}, X_i\right]- E_{GB}\left[\left(n-1\right)\mathrm{Fe}, Y_j\right] + E_{GB}\ ,\label{eq:int_GB}\\
    E_{int}^{bulk} = E_{bulk}\left[\left(\del{n}\add{m}-2\right)\mathrm{Fe}, X_i, Y_j\right] - E_{bulk}\left[\left(\del{n}\add{m}-1\right)\mathrm{Fe}, X\right]- E_{bulk}\left[\left(\del{n}\add{m}-1\right)\mathrm{Fe}, Y\right] + E_{bulk}\ .\label{eq:int_bulk}
\end{gather}

A positive interaction energy between two solutes indicates repulsion, while a negative value indicates attraction. Hence, the arrangement of two solutes in the structure is energetically favorable in the latter case. We further stress the explicit dependence of the $E_{int}$ on particular occupied sites $i$, $j$, which, in the case of bulk, is effectively reduced to the distance (coordination shell) between the two species sites $i$ and $j$. Finally, since all sites in bulk are equivalent, the site indices in single-solute terms in Eq.~\eqref{eq:int_bulk} diminish.

\subsection{Mechanical properties} 

\subsubsection{Impact on cohesion}
The mechanical strength of an interface is, based on the Rice–Thomson–Wang model, controlled by the work of separation, $W_{sep}$~\cite{rice1974ductile, rice1989embrittlement}. To determine this thermodynamic quantity, the GB structure is separated at a specific plane, resulting in a model with two additional free surfaces. The parameter $W_{sep}$ can then be evaluated using:
\begin{equation}
    W_{sep} = \frac{E_{GB}^{sep} - E_{GB}}{A}
\end{equation}
Here, $E_{GB}^{sep}$ and $E_{GB }$ represent the energy of the separated and unseparated GB structures, respectively. Previous studies have outlined two computational approaches for computing $W_{sep}$. The first method yields the rigid work of separation, denoted as $W_{sep}^{RGS}$, which is obtained without the relaxation of the newly created surfaces. The second method incorporates surface relaxation and is more traditionally used. In the present study, we exclusively employed the former approach, i.e., $W_{sep} \equiv W_{sep}^{RGS}$, to minimize additional computational expenses. All planes parallel to the GB were examined to identify the plane with the lowest cleavage energy.

The change in cohesion, denoted as $\eta(X)$, characterizes the effect of a solute $X$ on the $W_{\text{sep}}$:
\begin{equation}
    \eta(X) = W_{sep}(X) - W_{sep}(\emptyset)
\end{equation}
Here, $W_{sep}(X)$ and $W_{sep}(\emptyset)$ represent the values of $W_{sep}$ for a GB with and without (i.e., pristine GB) a segregant $X$, respectively. A negative value indicates that the segregant induces embrittlement of the GB, while a positive value suggests a cohesion-enhancing effect. Notably, $W_{sep}(X)$ was exclusively computed with respect to the minimum segregation energy site to mitigate computational costs. Since we calculate $W_{sep}$ without considering surface relaxations, we will refer to the resulting cohesion change as $\eta_{RGS}$ from this point onward.

\subsection{Thermodynamic modeling}
The equilibrium enrichment is modeled using an extension of the McLean formalism~\cite{mclean1958grain, lejvcek2010grain}. This model incorporates the segregation of multiple components to multiple GB sites and can be calculated using the following equation:
\begin{equation}
    c_{GB}(X) = \frac{1}{N}\sum_k \frac{c_{bulk}(X) \exp\left(- \frac{E_{seg}^{k}(X)}{k_BT}\right)}{1 + \displaystyle\sum_{Y} c_{bulk}(Y) \left[\exp\left(-\frac{E_{seg}^{k}(Y)}{k_BT}\right) - 1\right]}\ .
\end{equation}
Here, the $X$ and $Y$ iterate over all components (solute atoms), and the index $k$ covers all distinct GB sites. The GB and bulk concentrations of solute $X$ are represented as $c_{GB}(X)$ and $c_{bulk}(X)$, respectively. The total number of GB sites is denoted as $N$, and the segregation energy of a solute $X$ to the GB site $k$ is expressed as $E_{seg}^{k}(X)$. \add{We note, that this model accounts for site completion for each species individually, but without considering solute--solute interactions.}

Given that segregation is a time-dependent process primarily controlled by diffusion phenomena, we have incorporated a kinetic model into our work, as described in Refs.~\cite{scheiber2018kinetics, scheiber2020solute}. It accounts for the multi-site nature of GBs and enables the prediction of GB occupation by multiple components over time during heat treatments. The model can capture site competition among solutes; however\add{, similar to the extension of the McLean equation}, it \add{currently} does not \del{currently} account for solute-solute interactions. Our python-based implementation of the kinetic model is publicly available on GitHub~\cite{kinetics_amin}. We apply this multi-site, multi-component kinetic model to investigate $\Sigma 3(1\bar{1}1)$ GB occupation for two different alloy compositions and three distinct isothermal temperatures known to induce RTE: specifically, at 650, 550, and 400$\,\degree\text{C}$. In the first case, we examine the GB enrichment of 0.05$\,\text{wt}.\%$ Sn (equivalent to 500 wt. ppm) in conjunction with 0.05$\,\text{wt}.\%$ Sb. In the second case, we replace Sb with 2$\,\text{wt}.\%$ Ni. \add{Although these concentrations may seem high in the light of conventional steel compositions, they are deliberately chosen so to highlight the enrichments. Additionally, the rise in recycling rates will lead to increased concentrations of the tramp elements in future steels.} We initiate our simulations at 2000\,K to allow for the equilibration of GB and bulk concentrations of the respective elements, with a chosen grain diameter of 100$\,\micro\text{m}$ and a GB thickness of 8.4$\,\AA$. \add{We note that this is an artificial temperature only serving to have fast enough kinetics but not representing any real state of the material. Thereby,} \del{At this temperature,} the solutes exhibit sufficient mobility to align with the McLean equation. Subsequently, the system undergoes cooling to ambient temperature following Newton's law of cooling, characterized by a cooling rate of $r = 0.1\,$s$^{-1/2}$. \add{Again, this step does not represent any real cooling process but rather allows for the site occupation redistribution limited by kinetics but reflecting site competition.} The GB concentrations obtained in this manner serve as a suitable initial point for further investigations of RTE effects and can simulate the cooling of steel during production, from high temperatures to room temperature. Alternatively, the McLean equation can be used to calculate equilibrium concentrations at 25$\,\degree\text{C}$, providing an upper limit. However, these thermodynamically derived concentrations tend to be high, resulting in GB depletion when heating to RTE critical temperatures. The material is then heated to 900$\,\degree\text{C}$ at a constant heating rate of 5$\,\text{K/s}$. After achieving equilibrium, the sample is quenched with a rate of 50$\,\text{K/s}$ to reach either 650, 550, or 400$\,\degree\text{C}$, where it is isothermally heated for $300\cdot10^6\,\text{s}$. \add{DFT-based diffusion data for the considered solutes, including the diffusion prefactor $D_0$ and activation coefficient $Q_A$ were taken from Ref.~\cite{versteylen2017first}, and are summarized in Supplementary Material Table~S8.}

\section{Results}
\subsection{Structure and properties of pristine grain boundaries}
Table~\ref{tab: GB_properties} provides an overview of the relaxed cell dimensions and the computed GB properties. All values are consistent with those reported in prior literature~\cite{bhattacharya2014ab, mai2022segregation, yang2023high, jiang2022effects}. Notably, among the three investigated GB types, the $\Sigma 3(1\bar{1}2)$ GB, in which local atomic environments closely resemble bulk, exhibits the lowest GB energy. This observation aligns with the general trend that higher GB energies are associated with greater interface distortions and will be important for further discussion of segregation behavior.

\begin{table}[tb!]
\centering
\caption{Summary of properties for the studied GBs, including the number of atoms in the slab ($N_{GB}$), dimensions of the simulation boxes ($b$ and $c$), corresponding GB energies ($\gamma_{GB}$), excess volume ($V_{exc}$), and work of separation ($W_{sep}$). Note that the $a$ dimension is constant across the investigated GB types and equals 4.015$\,\AA$; it has been omitted from the table for clarity. Additionally, the $c$ dimension includes a vacuum layer on top of the GB slab, with a converged thickness of at least $6\,\AA$.}
\label{tab: GB_properties}
\begin{threeparttable}

\begin{tabular}{c c c c c c c c}
 \hline
 System & $N_{GB}$ & $b$ ($\AA$) & $c$ ($\AA$) & $\gamma_{GB}$ (J/m$^2$) & $V_{exc}$ ($\AA$) & $W_{sep}$ (J/m$^2$)\\
  \hline
 $\Sigma 3(1\bar{1}1)$ & 72 & 6.954 & 35.504 &  1.60 & 0.27 & 3.82\\
 & & & & 1.61\textcolor{blue}{\textsuperscript{a}}, 1.58\textcolor{blue}{\textsuperscript{b}} & 0.31\textcolor{blue}{\textsuperscript{a}}, 0.301\textcolor{blue}{\textsuperscript{b}} & 3.81\textcolor{blue}{\textsuperscript{b}}, 3.885\textcolor{blue}{\textsuperscript{c}}, 3.825\textcolor{blue}{\textsuperscript{d}}\\
 $\Sigma 3(1\bar{1}2)$ & 48 & 4.917 & 33.816 &  0.42 & 0.11 & 4.74\\
  & & & & 0.43\textcolor{blue}{\textsuperscript{a}}, 0.45\textcolor{blue}{\textsuperscript{b}} & 0.10\textcolor{blue}{\textsuperscript{a}}, 0.124\textcolor{blue}{\textsuperscript{b}} & 4.72\textcolor{blue}{\textsuperscript{b}}\\
 $\Sigma 9(2\bar{2}1)$ & 70 & 6.348 & 40.068 &  1.72 & 0.25 & 3.56\\
 & & & & 1.71\textcolor{blue}{\textsuperscript{a}}, 1.75\textcolor{blue}{\textsuperscript{b}} & 0.26\textcolor{blue}{\textsuperscript{a}}, 0.279\textcolor{blue}{\textsuperscript{b}} & 3.60\textcolor{blue}{\textsuperscript{b}}\\
 \hline
\end{tabular}
\begin{tablenotes}
    \item[\textcolor{blue}{a}] PAW-GGA~\cite{bhattacharya2014ab}.
    \item[\textcolor{blue}{b}] PAW-GGA~\cite{mai2022segregation}.
    \item[\textcolor{blue}{c}] PAW-GGA~\cite{yang2023high}.
    \item[\textcolor{blue}{d}] PAW-GGA~\cite{jiang2022effects}.
  \end{tablenotes}
\end{threeparttable}
\end{table}

\subsection{Segregation energies of single species and impact on cohesion}
Solute segregation energies are plotted as functions of GB distance in Fig.~\ref{fig:seg_profiles}, while the resulting magnetic moments of segregants versus distance are shown in Supplementary Material Fig.~S2. Across all GB types, the tramp elements As, Sb, and Sn consistently exhibit stronger segregation binding compared to the alloying elements Ni, Cr, and Mo. This is not only for the minimum energy segregation site but also for most other sites near the GB, as evident from Fig.~\ref{fig:seg_profiles}. A closer examination reveals that the atoms Mo, Cr, Sb, and Sn favor the $\Sigma 3(1\bar{1}1)$ GB plane, while Ni and As prefer the site adjacent to the GB plane (Fig.~\ref{fig:seg_profiles}a). A similar trend is observed in the $\Sigma 9(2\bar{2}1)$, where Mo, Sb, and Sn exhibit the highest segregation tendency to site number 3 (located at the GB plane, Fig.~\ref{fig:seg_profiles}c), while Ni and As favor site 1. However, while the site with the second-highest segregation tendency for Sn and Sb remains at the GB plane (labeled as site 2), As and Ni show a greater segregation tendency for sites closest to the GB plane (labeled as site 4). In summary, the segregation profile trends in the $\Sigma 3(1\bar{1}1)$ and $\Sigma 9(2\bar{2}1)$ GBs can be grouped based on similarity with Mo, Sb, and Sn in one group and Ni and As in another group. 

\add{The segregation energies are tabulated in Supplementary data (Tables S3--S5) along with the Voronoi volumes of individual segregation sites. Although a general correlation between $E_{seg}$ and the Voronoi volume exists, it is not a rule that the largest segregation tendency is associated with the largest void. In particular, As seems to be an outlier suggesting that a large portion of $E_{seg}$ stems from chemical bonding effects.} \add{In addition, we investigated the segregation tendency of Sn, Sb, and As to interstitial GB sites, as proposed by Lejcek $\etal$~\cite{lejvcek2021entropy}, who suggested that these elements segregate to both substitutional and interstitial sites. However, in the GB models investigated here, we found that interstitial segregation is only possible to the central void located at the $\Sigma 3(1\bar{1}1)$ GB, with $E_{\text{seg}}^{Sn} = -0.26\, \text{eV}$, $E_{\text{seg}}^{Sb} = -0.27\, \text{eV}$, and $E_{\text{seg}}^{As} = -0.69\, \text{eV}$ and, thereby, interstitial segregation remains less favorable than substitutional segregation, which exhibits $E_{\text{seg}}^{Sn} = -1.30\, \text{eV}$, $E_{\text{seg}}^{Sb} = -1.24\, \text{eV}$, and $E_{\text{seg}}^{As} = -0.87\, \text{eV}$. Overall, our findings suggest that substitutional segregation predominates over interstitial segregation in the studied GBs and, consequently, was not considered in the co-segregation studies.} \add{The used simulation models are relatively small, therefore we performed additional calculations to investigate possible size effects. Namely, we used the $\Sigma 3(1\Bar{1}2)$ GB (48 atoms), and created a larger 2$\times$2$\times$1 supercell consisting of 192 atoms. Considering Sn, which exhibits the strongest segregation binding among all solutes, the predicted segregation energy decreased by $0.03\uu{eV}$, from $-0.33\,$eV in the 48-atom cell to $-0.36\,$eV in the larger supercell. This comparison demonstrates that the size effects are at least an order of magnitude smaller than the reported segregation behavior, and the chosen system sizes are adequate.} \add{Another peculiarity is the slow convergence of $E_{\text{seg}}$ for Sn and Sb in the $\Sigma 3(1\bar{1}1)$ GB, and Cr and Mo in the $\Sigma 3(1\bar{1}2)$ GB with distance from the GB. We have also tested sites farther away from the GB, and the $E_{seg}$ kept decreasing. We, therefore, ascribe this behavior to a strong elastic interaction between those large elements and the GB.}
 
\begin{figure}[htp]
		\centering
		    {\footnotesize(a)\qquad $\Sigma 3(1\bar{1}1)[110]$}\\
    		\includegraphics[width=8cm]{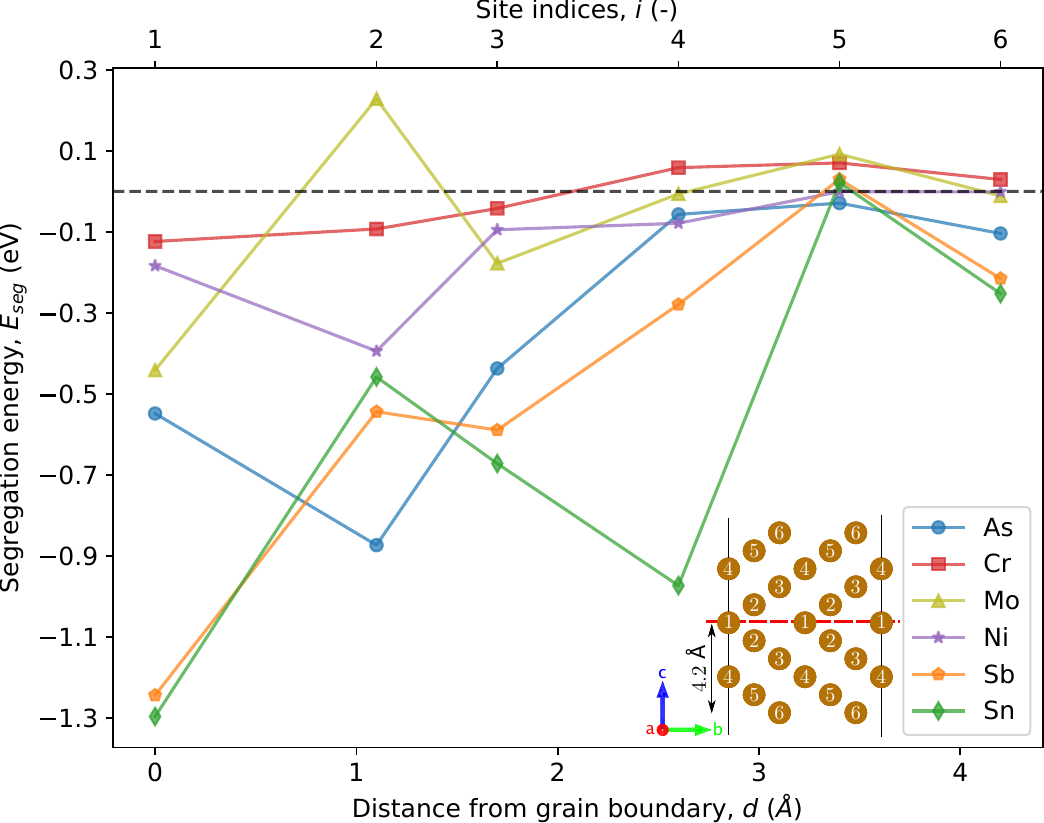}\smallskip\\
    		{\footnotesize(b)\qquad $\Sigma 3(1\bar{1}2)[110]$}\\
    	    \includegraphics[width=8cm]{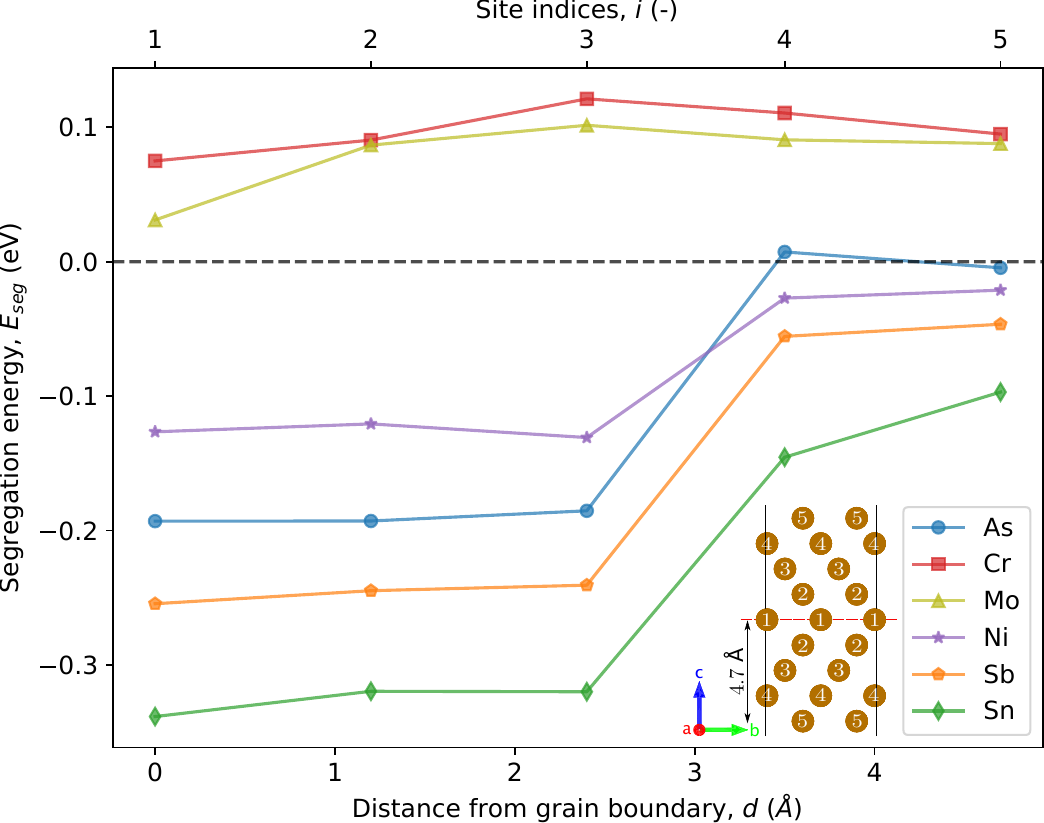}\smallskip\\
            {\footnotesize(c)\qquad $\Sigma 9(2\bar{2}1)[110]$}\\
    	    \includegraphics[width=8cm]{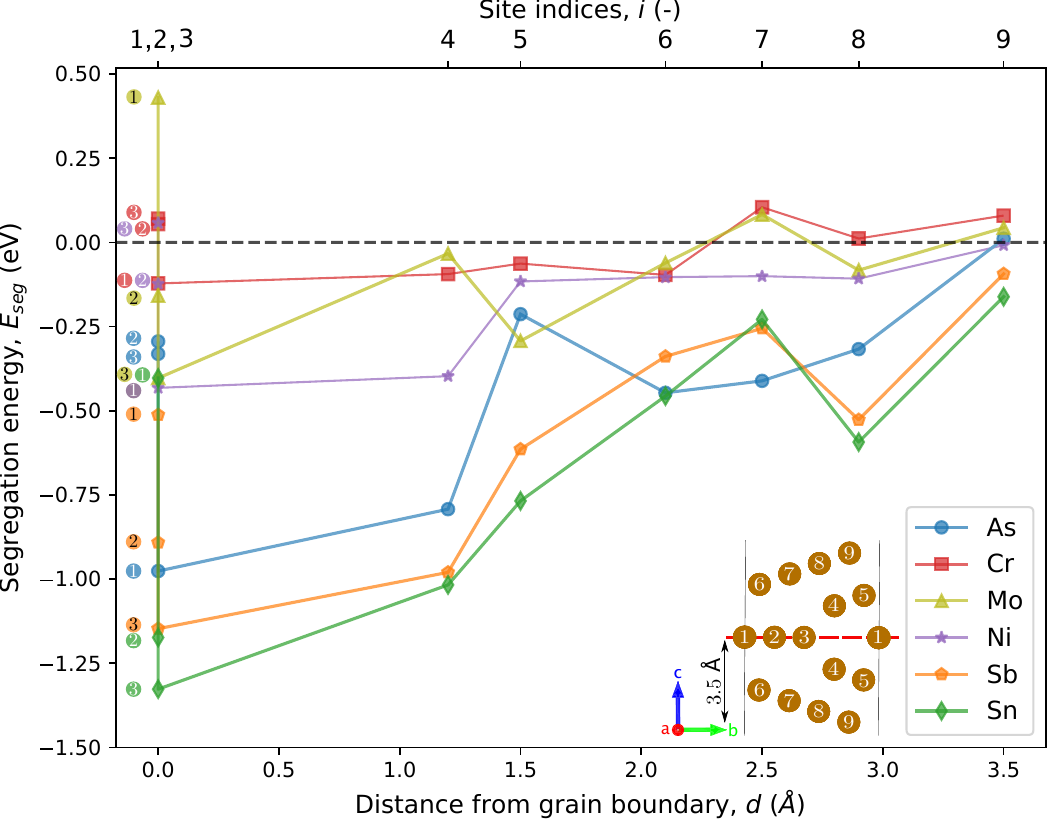}
		\caption{Segregation energy of a single solute plotted against the distance from GB in (a) $\Sigma 3(1\bar{1}1)$, (b) $\Sigma 3(1\bar{1}2)$, and (c) $\Sigma 9(2\bar{2}1)$ grain boundaries. The upper x-axis refers to the indexing of sites used in the text.}
		\label{fig:seg_profiles}
\end{figure}

\begin{figure}[htb!]
		\centering
		    {\footnotesize(a)}\\
    		\includegraphics[width=8cm]{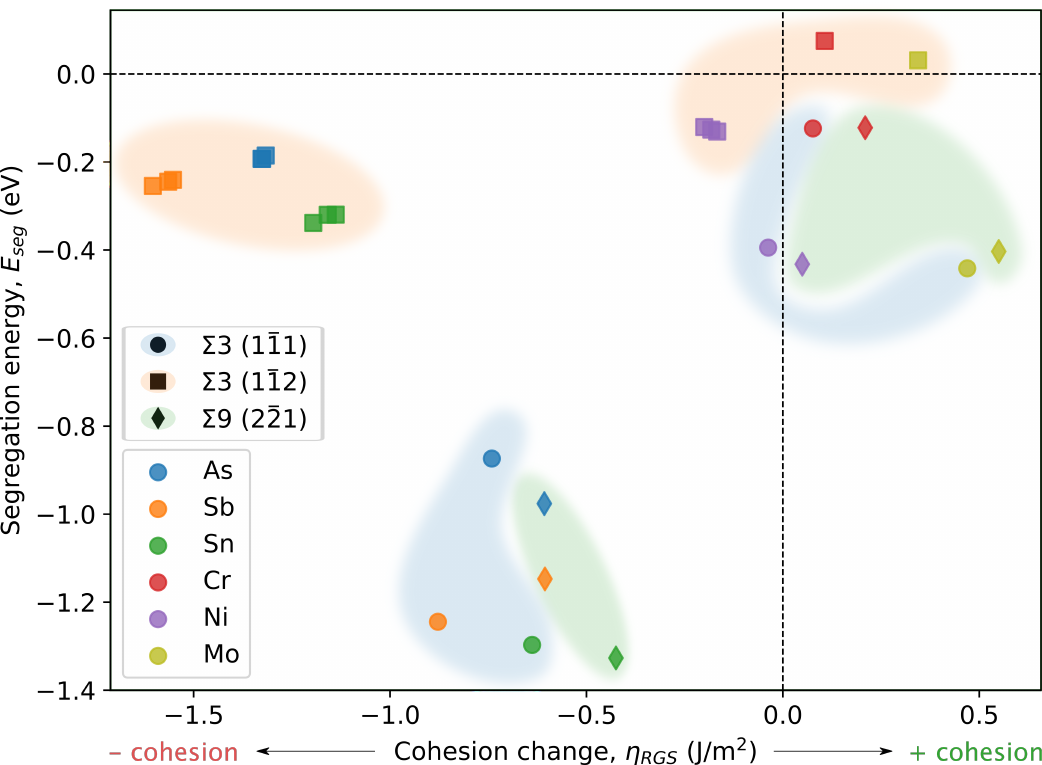}\\
    		{\footnotesize(b)}\\
    	    \includegraphics[width=8cm]{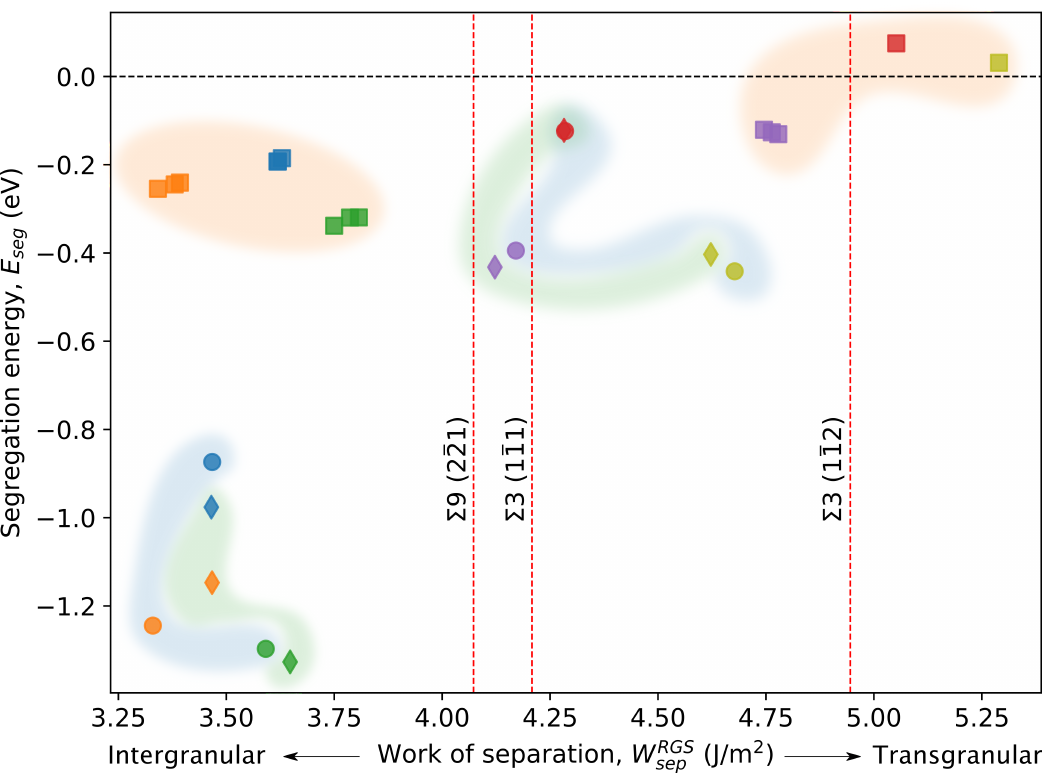}
		\caption{Impact of single solute segregation on: (a) GB cohesion, $\eta_{RGS}$ (J/m$^2$), (b) work of separation, $W_{sep}^{RGS}$ (J/m$^2$)\add{, lowering this value increases the tendency for intergranular fracture, as it weakens the GB cohesion}. The red dotted lines indicate the rigid work of separation of the pure GB.}
		\label{fig:single_segreg_mech_properties}
\end{figure}

The minimum segregation energies of all elements for the three considered GBs are depicted in Fig.~\ref{fig:single_segreg_mech_properties}, illustrating their effects on the rigid work of separation. Since the $\Sigma 3(1\bar{1}2)$ GB exhibited multiple sites (1, 2, and 3 in Fig.~\ref{fig:seg_profiles}b) with segregation energy differences less than $0.05\,$eV, we included all in the graph, too. Comparative analyses with literature data from DFT and experimental sources are provided in Supplementary Material Tables~S1 and S2.     

Clearly, the degree of segregation significantly hinges on the specific GB type. The $\Sigma 3(1\bar{1}2)$ GB, characterized by a bulk-like structure and consequently possessing a smaller excess volume relative to the other two GBs, exhibits reduced segregation tendencies (w.r.t. the other two considered GBs) for all investigated elements. For the tramp elements (As, Sb, Sn), this difference amounts to approximately $0.6$ to $1.0\,$eV.

A notable observation is the absence of segregation tendencies for Cr and Mo at the $\Sigma 3(1\bar{1}2)$ GB. This contradicts the previously predicted~\cite{mai2022segregation} minor to negligible segregation of these elements (as indicated in Table~S1). Given that the work of Mai et~al.~\cite{mai2022segregation} appears to be the only other investigation of this element--GB combination, statistical evidence is lacking, and the discrepancy of less than $0.17\,$eV could potentially be attributed to variations in calculation parameters.

The segregation energies in the $\Sigma 3(1\bar{1}1)$ and $\Sigma 9(2\bar{2}1)$ GBs are closely aligned, in particular for the alloying elements Cr, Ni, and Mo; the maximum disparity of $0.11\,$eV is obtained in the case of As. In a broader context, the trend of segregation potency follows the sequence Sn$>$Sb$>$As$>$Mo$\approx$Ni$>$Cr.

A clear embrittling effect becomes evident for the tramp elements after examining the influence of single-element segregation on the work of separation. Notably, this effect is nearly twice as pronounced in the bulk-like GB compared to the other GBs. This distinction can be attributed to the smaller excess volume characteristic of the $\Sigma 3(1\bar{1}2)$ GB, resulting in more substantial local distortions stemming from segregation. Consistent with prior investigations~\cite{mai2022segregation, geng2000effect}, the impact of Cr and Mo on the GBs is opposite, as both contribute to the reinforcement of the GBs, with Mo being approximately twice stronger than Cr. Furthermore, the effect of Ni on the GBs displays a dependence on the GB type. Specifically, while Ni weakens the $\Sigma 3(1\bar{1}2)$ and $\Sigma 3(1\bar{1}1)$ GBs, it exhibits a slight strengthening effect on the $\Sigma 9(2\bar{2}1)$ GB. Finally, we note that based on the magnitude of $W_{sep}$, all tramp elements are predicted to lead to intergranular fracture, aligning with experimental observations of RTE~\cite{Olefjord1978-df}. \add{This conclusion is based on Sutton's criterion, comparing work of separation at the GB with creating pristine surfaces by cleaving the bulk. Based on the DFT-based surface energies~\cite{Tran2016-uv}, the lowest surface energy of bcc-Fe is for $(110)$ surface and has value $2.45\uu{J/m^2}$, suggesting that $W_{sep}\lessapprox4.9\uu{J/m^2}$ should prefer fracture along GB rather than through the grain interior.}

\subsection{Co-segregation energetics}\label{cosegs_ints}
We analyzed and consolidated the incremental segregation energies ($E_{seg}^{inc}$) for all elemental co-segregation combinations, presenting the results in heatmap format within Table~\ref{tab:incremental_co_seg}. Furthermore, the interaction energies for the co-segregated pairs' minimum energy configurations at the GBs are provided in Table~\ref{tab:interactions_co_seg}. 
From the 108 elemental combinations, 30 exhibit unfavorable incremental co-segregation, marked by vacant cells in Table~\ref{tab:incremental_co_seg}. A majority of these, namely 22 combinations, are for the $\Sigma 3(1\bar{1}2)$ GB. This behavior can be attributed to the bulk-like structure of the GB, coupled with lower individual segregation energies. Notably, Ni stands out in this trend: pre-segregated Ni to this GB amplifies the segregation propensity of the tramp elements. Similarly, Ni is more prone to segregate to $\Sigma 3(1\bar{1}2)$ when As, Sb or Sn are already present. We attribute this behavior to the robust, attractive interactions between Ni and tramp elements (cf. Table~\ref{tab:interactions_co_seg}). These interactions represent the most potent attractive forces in all studied scenarios.

The remaining instances of unfavorable co-segregation are observed in the $\Sigma 9(2\bar{2}1)$ GB. A closer examination of interatomic interactions (cf. Table~\ref{tab:interactions_co_seg}) within this GB reveals that almost all interactions are repulsive in nature, barring the negligible attraction of $-0.02\,$eV between Ni-Ni. Consequently, all co-segregation energies are less negative than the single-segregation scenario (corresponding cells in Table \ref{tab:incremental_co_seg} have blueish colors). This phenomenon is most evident in the case of Cr, where the modest segregation tendency of $E_{seg}(\mathrm{Cr})=-0.12\,$eV in solitary segregation diminishes when prior-segregated atoms impede the co-segregation of Cr. Mo exhibits the same behavior when co-segregating in the presence of pre-segregated As. However, the unfavorable co-segregation of Mo with Sb or Sn arises mostly from site preferences, since the repulsive interactions approx. $0.1\,$eV are rather small compared to the segregation energy of $-0.40\,$eV for Mo. As previously described Mo, Sb, and Sn share very similar segregation profiles across the studied GBs, and have the highest segregation tendency to the same sites. Consequently, Mo must occupy a less energetically favorable site at the GB, leading to its unfavorable co-segregation. 

The $\Sigma 3(1\bar{1}1)$ GB, does not exhibit any unfavorable co-segregation scenarios. However, it is noteworthy that in cases where repulsive interactions are present (e.g., As--Sb), the tendency for co-segregation is diminished relative to the single segregation scenario. The presence of attractive interactions between Sn and Sb pairs, spanning from $-0.10$ (Sn--Sn) to $-0.26\,$eV (Sb--Sb), enhances the likelihood of co-segregation for these pairs beyond that of individual segregation. Otherwise, the interactions occurring in the $\Sigma 3(1\bar{1}1)$ GB range from negligible attraction ($-0.08\,$eV $\leq E_{seg}^{int} \leq -0.02\,$eV) to significant repulsion ($0.21\,$eV $\leq E_{seg}^{int} \leq 0.37\,$eV) in the order As-As, As-Mo, As-Sn, As-Sb. 

It is important to note that the sequence of segregation events can also influence these interactions. Initially segregated As induces a repulsive force on Sb and Sn, while the interaction dynamics shift when Sb or Sn segregate first, leading to a more favorable, albeit modest, attractive force (cf. Table~\ref{tab:interactions_co_seg}). \add{The fact that the interaction energy is not symmetrical is caused by small structural changes induced upon the segregation of the first solute. Consequently, the modified local environment for the segregation of the second solute may result in an overall different local minimum on the Born-Oppenheimer potential energy surface, thus leading to the interaction energy being dependent on the segregation sequence.} \del{The fact that the interaction energy is not symmetrical is caused by the possible site competition in the case of the same minimum segregation energy site (Eq.~\eqref{eq:int_GB}).} A similar behavior can be observed in the case of co-segregation of Mo with Sb or Sn in the $\Sigma 9(2\bar{2}1)$ GB. Initial segregation of Mo results in strong repulsive forces on Sn and Sb (0.62$\,$eV and 0.49$\,$eV), while the repulsive interactions are reduced to 0.10$\,$eV and 0.09$\,$eV when the sequence of segregation is altered.

\begin{table*}[htb!]
    \centering
    \caption{The incremental segregation energy ($E_{seg}^{inc}(Y|X)$\add{, Eq.~\eqref{eq:E_seg^inc}}) of solute $Y$ (rows) when solute $X$ (columns) already occupies its most preferable site at the GB. The table is organized into three rows for each species corresponding to the $\Sigma 3(1\bar{1}1)$, $\Sigma 3(1\bar{1}2)$, and $\Sigma 9(2\bar{2}1)$ GBs. An absence of data within a cell indicates unfavorable incremental segregation ($E_{seg}^{inc}(Y|X)\geq0$). The heatmap further shows the difference between the $E_{seg}^{inc}(Y|X)$ and the single segregation energy ($E_{seg}(Y)$) of solute $Y$.}
    \includegraphics[width=13cm]{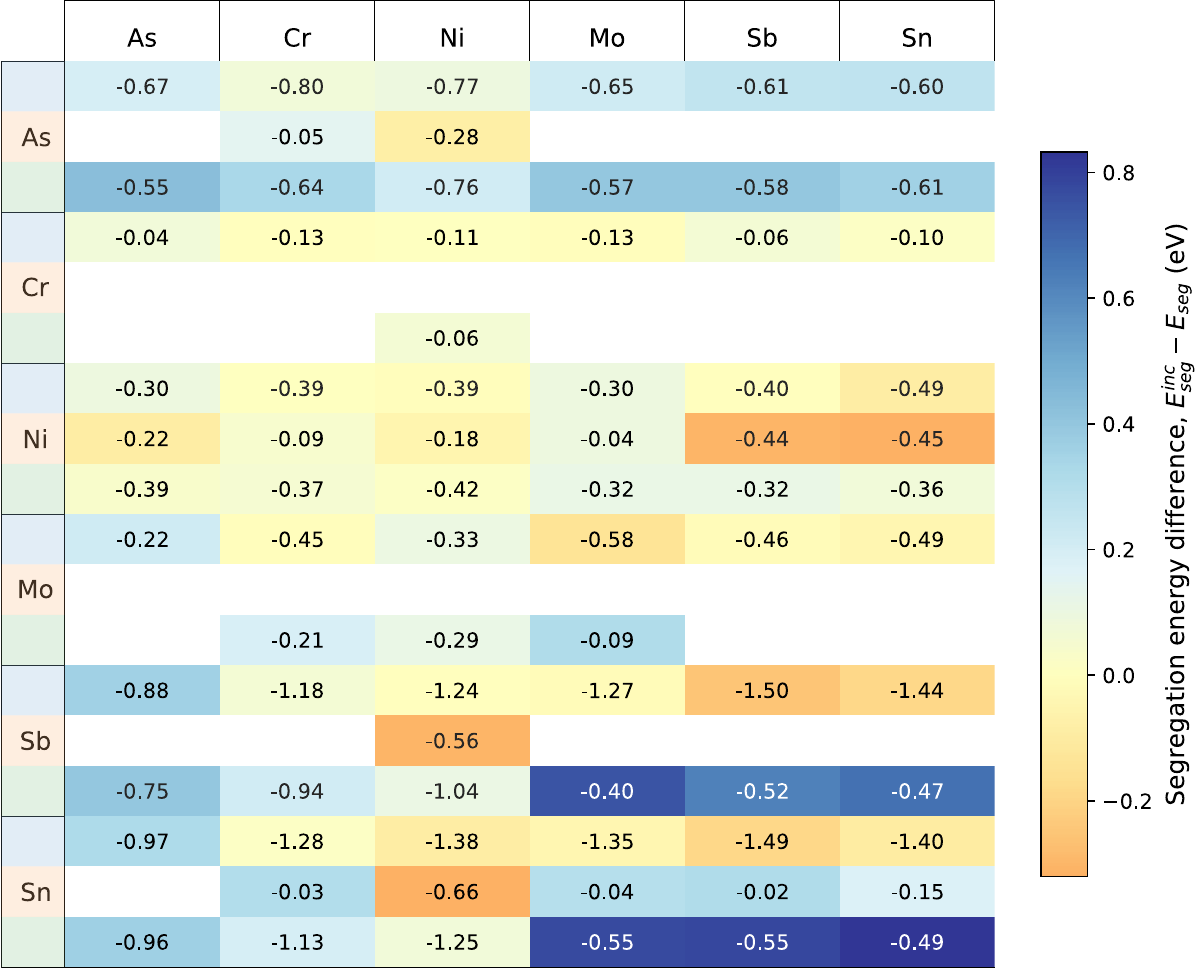}
    \label{tab:incremental_co_seg}
\end{table*}

\begin{table}[htb!]
    \centering
    \caption{The interactions at GB ($E_{int}^{GB}$\add{, Eq.~\eqref{eq:int_GB}}) involving various elemental combinations. Solute 1 (columns) is positioned in its minimum energy configuration, while solute 2 (rows) is systematically adjusted around it until the minimum energy state is attained. This could result in non-symmetrical interaction energies in the case of the same minimum segregation energy site (site competition). The three rows for each species correspond to the $\Sigma 3(1\bar{1}1)$, $\Sigma 3(1\bar{1}2)$, and $\Sigma 9(2\bar{2}1)$ GBs.}
    \includegraphics[width=13cm]{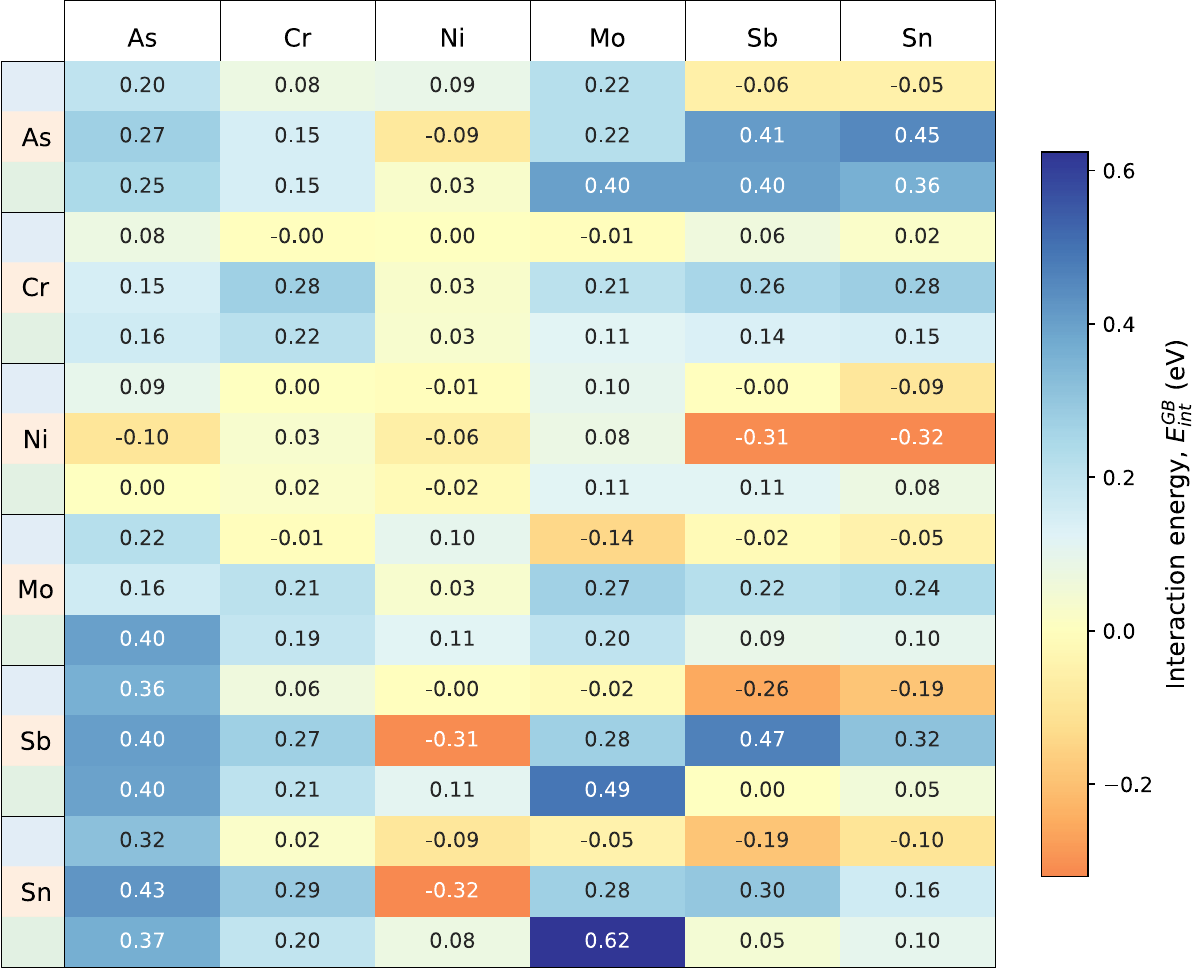}
    \label{tab:interactions_co_seg}
\end{table}

To delve deeper into the segregation behavior and check if potential attractive bulk interactions ($E_{int}^{bulk}$) could hinder segregation, we broaden our perspective to include these as well. Assumably, these interactions can also significantly differ from those observed at the GB. Considering the limited available literature on bulk interactions concerning tramp elements, we extended our investigation to cover the pair interactions within the first four nearest neighbor (nn) distances. The results are consolidated in Table~\ref{tab:final_bulk_interactions}. The interactions obtained between Cr, Ni, and Mo align well with those reported in Ref.~\cite{gorbatov2016first}.

The bulk interactions from the third shell on are small, ranging from $-0.07\,$eV $\leq E_{int}^{bulk}\leq0.10\,$eV. All other interactions occurring on the first two nn-distances are mostly repulsive, with the exception of Ni--As, Ni--Sb, and Ni--Sn, where the interactions are slightly attractive for the first nn-distance ($-0.05$, $-0.07$, and $-0.09\,$eV) and change to repulsive for the second-nearest neighbor distance ($0.12$, $0.13$, and $0.11\,$eV). The strength of repulsive interactions can be sorted as Sb--Sb $>$ Sb--Sn $>$ As--Sb $\approx$ Sn--Sn $>$ As--Sn $\approx$ As--As. This is followed by repulsive interactions within the range of $0.16\,$eV $\leq E_{int}^{bulk}\leq0.32\,$eV for Mo/tramp element combinations and in the range of $0.10\,$eV $\leq E_{int}^{bulk}\leq0.22\,$eV for Cr/tramp element combinations. In summary, the calculated interactions suggest that all the investigated elements tend to increase the distance between them and, hence, do not form intermediate phases in bulk Fe.

\begin{table}[htb!]
    \centering
    \caption{Pair interactions ($E_{int}^{bulk}$\add{, Eq.~\eqref{eq:int_GB}}) of all solute combinations in the bulk phase. Four rows representing the solute interactions at the four nearest neighbor distances (2.45, 2.83, 4.01, and 4.70 $\AA$) in FM bcc Fe are given for each species.}
    \includegraphics[width=13cm]{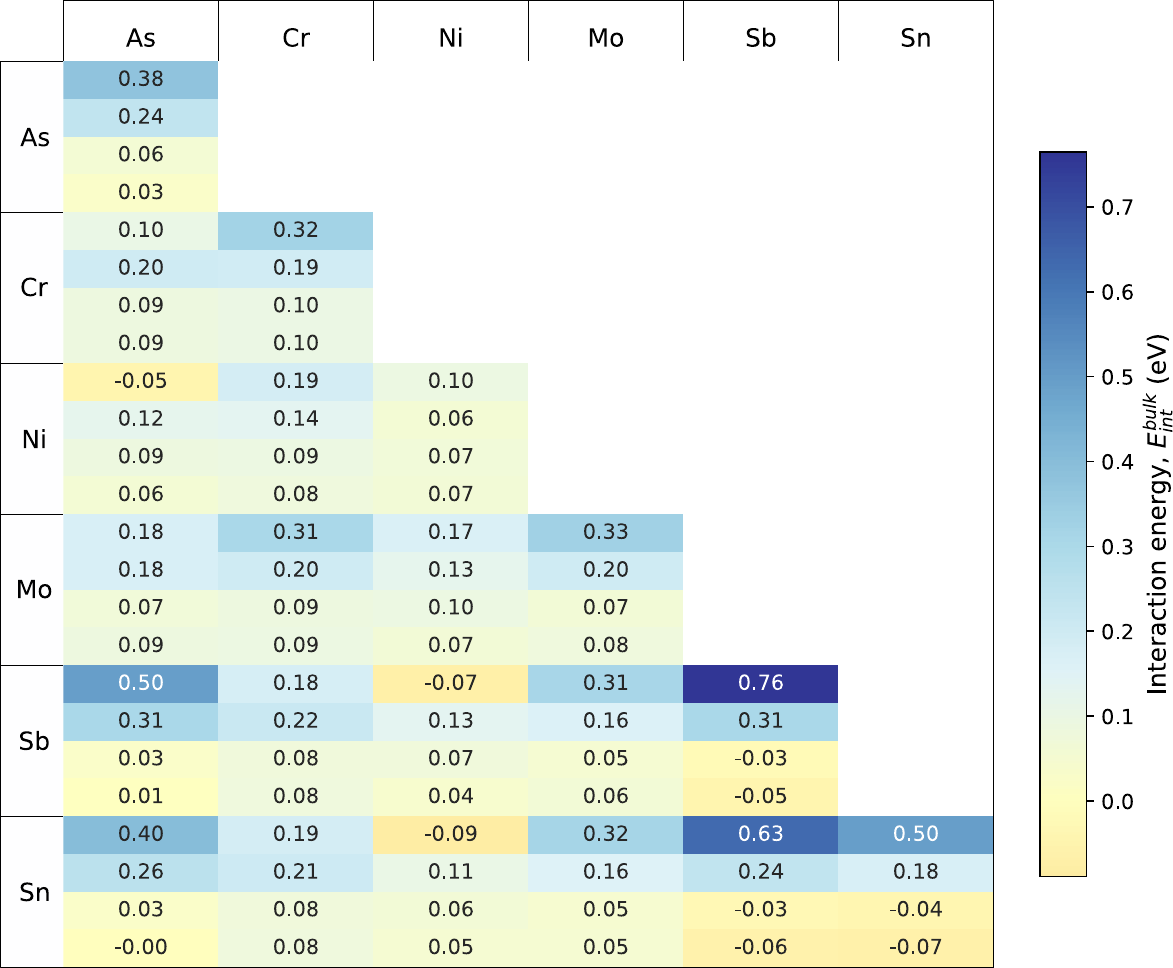}
    \label{tab:final_bulk_interactions}
\end{table}

\subsection{Impact of co-segregation on cohesion}
\begin{figure}[htp]
		\centering
		    {\footnotesize(a)\qquad $\Sigma 3(1\bar{1}1)[110]$}\\
    		\includegraphics[width=8cm]{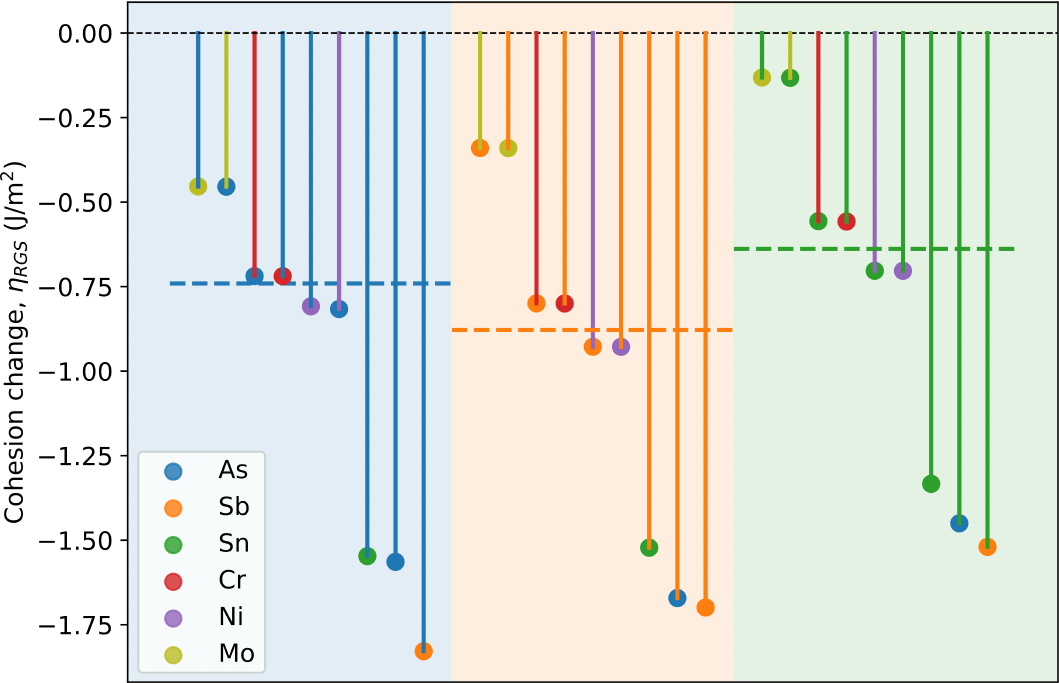}\smallskip\\
    		{\footnotesize(b)}\qquad $\Sigma 3(1\bar{1}2)[110]$\\
    	    \includegraphics[width=8cm]{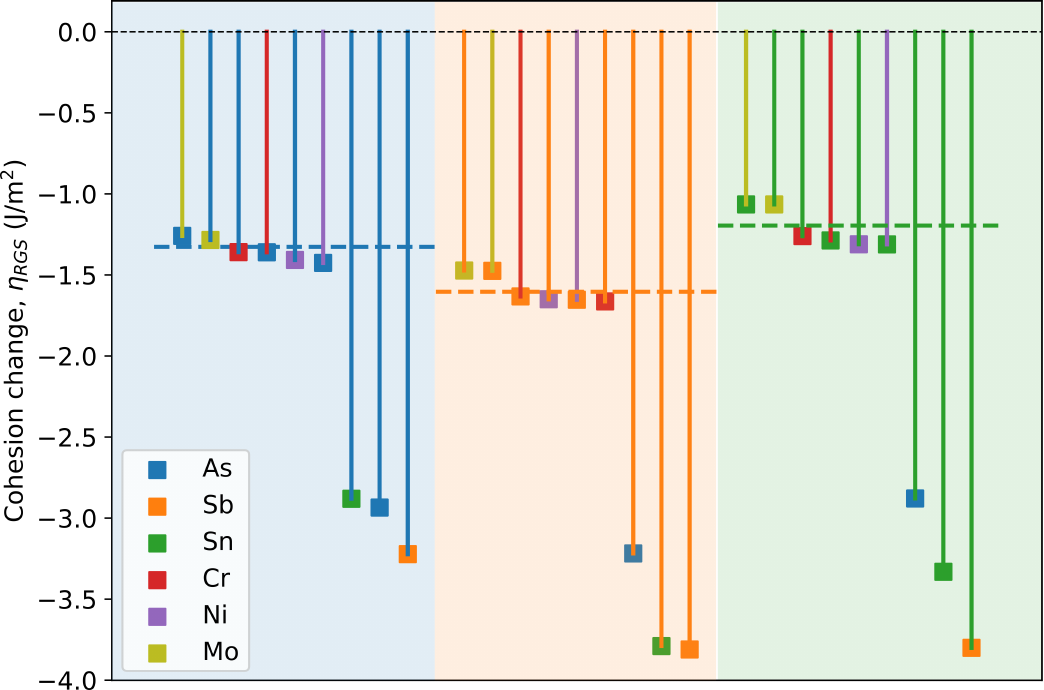}\smallskip\\
            {\footnotesize(c)\qquad $\Sigma 9(2\bar{2}1)[110]$}\\
    	    \includegraphics[width=8cm]{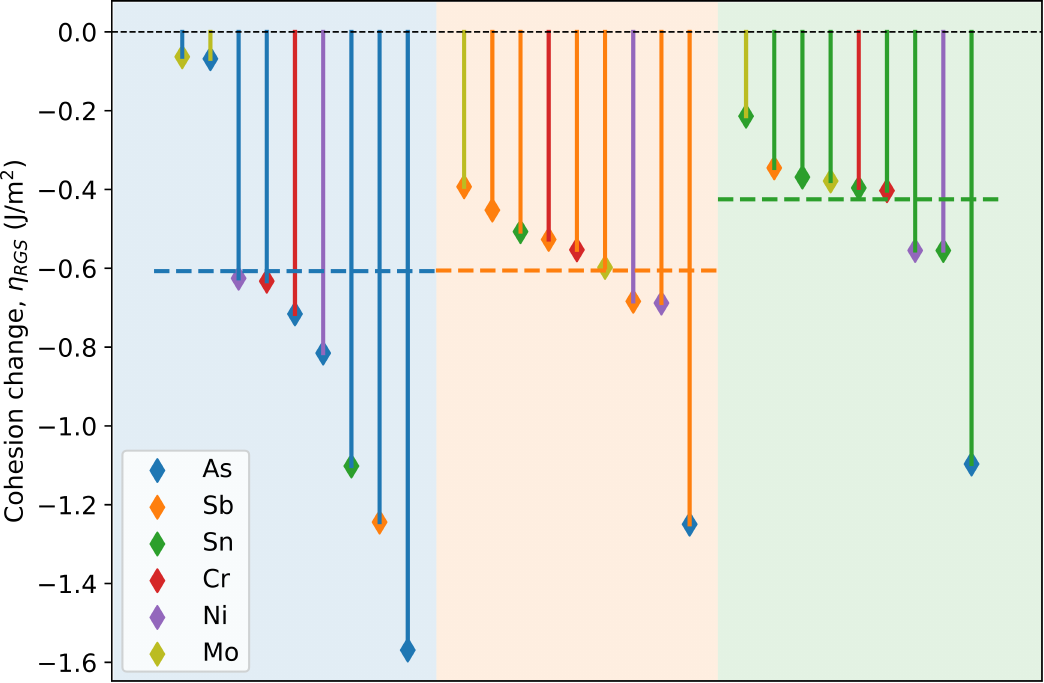}
		\caption{Impact of co-segregation on the GB cohesion, $\eta_{RGS}$ (J/m$^2$), in the (a) $\Sigma 3(1\bar{1}1)$, (b) $\Sigma 3(1\bar{1}2)$, and (c) $\Sigma 9(2\bar{2}1)$ GBs. Line colors represent the first segregating species, while symbol colors represent the co-segregated species. The single-element cohesion effects of As, Sb, or Sn are indicated by the colored dashed lines for reference.}
		\label{fig:multi_coseg_mech_properties}
\end{figure}

The influence of co-segregated solute pairs on GB cohesion is exemplified in Fig.~\ref{fig:multi_coseg_mech_properties}. We have deliberately focused our analysis on combinations involving As, Sb, or Sn as either the primary or secondary segregating element, excluding other combinations for the sake of graph clarity. To allow for easy comparisons with the sole elemental effects of As, Sb, or Sn on cohesion, dashed horizontal lines are included in the graphs as references. We remind the reader that our predicted effects on GB cohesion align with the energy-minimized configurations within the rigid work of separation. We also note that the negative values of $\eta_{RGS}$ represent embrittling w.r.t. pure FM bcc Fe.

Let us begin by investigating the influence of co-segregated Mo in conjunction with one of the tramp elements. A consistent trend emerges across all GB types: Mo mitigates the adverse effects of As, Sb, and Sn. This positive impact is particularly pronounced in the $\Sigma 3(1\bar{1}1)$ and $\Sigma 9(2\bar{2}1)$ GBs, while its influence in the $\Sigma 3(1\bar{1}2)$ is negligible. 

Turning our attention to Cr, its impact on cohesion varies. It enhances cohesion in the $\Sigma 3(1\bar{1}1)$ GB, embrittles the $\Sigma 3(1\bar{1}2)$, and exhibits dependency on the tramp element for the $\Sigma 9(2\bar{2}1)$, leading to cohesion reduction in combinations with As and enhancement in combinations with Sb and Sn. Finally, Ni consistently lowers the GB cohesion across all GB types and elemental combinations. Its impact, however, is not strong, except in the $\Sigma 9(2\bar{2}1)$, where it exerts a slightly more pronounced embrittling. It is noteworthy that the cumulative impact of Cr/Ni--As/Sb/Sn combinations mirrors the order of elemental impacts of the tramp elements, continuing to exert a detrimental effect on overall GB cohesion. The most substantial deleterious effects are observed in combinations of As, Sb, and Sn, resulting in approximately twice the magnitude of impact compared to individual elemental cases. Intriguingly, Sn--Sn, Sn--Sb, and Sb--Sb combinations within the $\Sigma 9(2\bar{2}1)$ appear to enhance cohesion compared to their individual counterparts, although they still ultimately lead to a decrease in GB cohesion w.r.t. pure Fe. This effect may be attributed to the preference of both elements for the same GB site directly at the GB plane. When one of these atoms occupies this site, the other must substitute a site further from the GB plane, which is energetically less favorable. It is also worth noting that within the $\Sigma 9(2\bar{2}1)$, combinations of As, Sb and Sn with As exhibit notably more pronounced detrimental effects.

\subsection{GB enrichment in the thermodynamic limit}
\begin{figure}[htp]
    \centering
    \includegraphics[width=8cm]{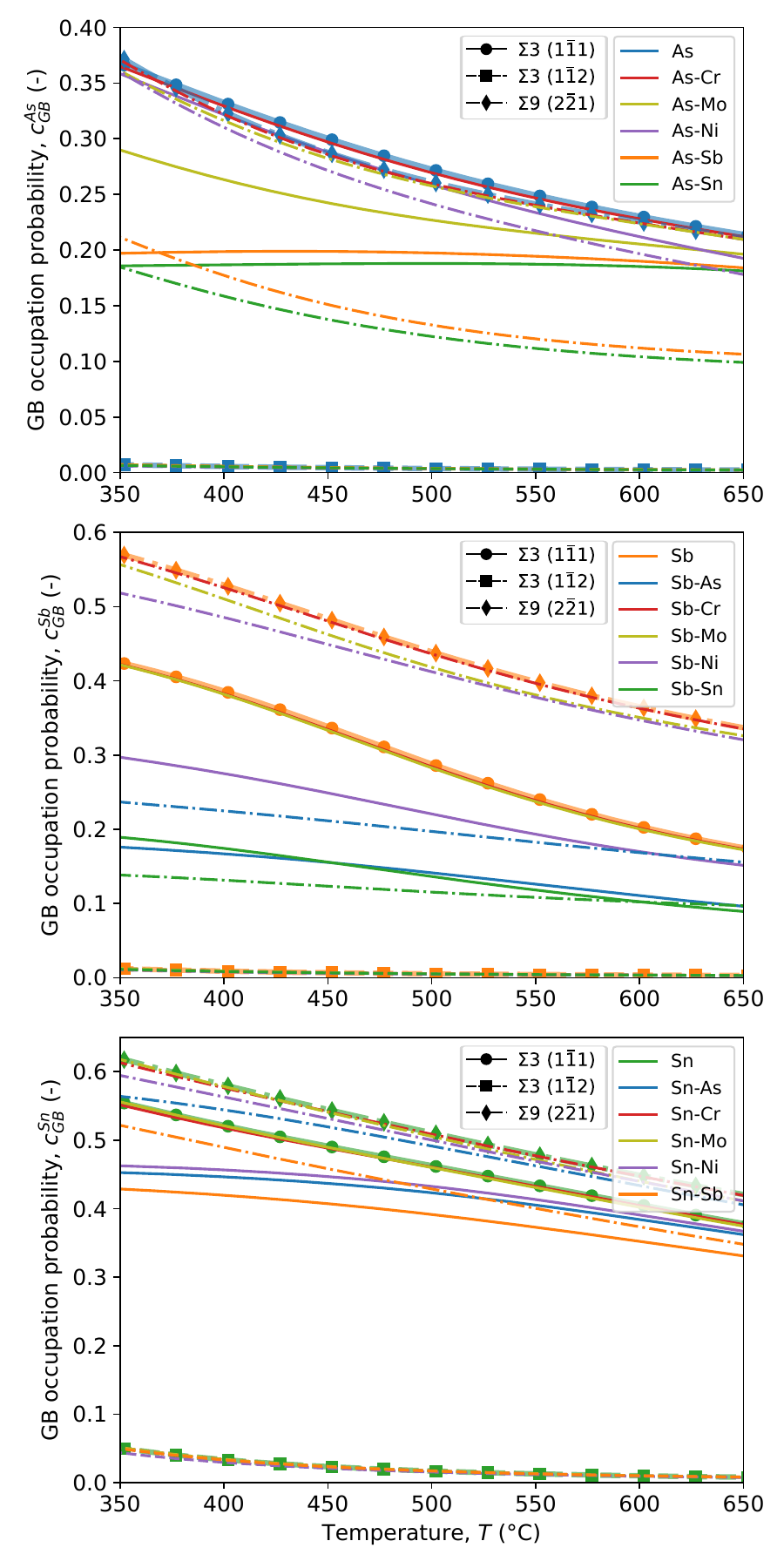}
    \caption{The GB enrichment of As (top), Sb (middle), and Sn (bottom) is depicted as a function of temperature and composition. The total concentration of As, Sb, and Sn remains constant at $0.05\,\textrm{wt.}\%$, while the alloying elements Cr, Mn, and Ni are consistently set at $2\,\textrm{wt.\%}$. The temperature range along the $x$-axis is chosen to represent the critical temperature region for RTE, spanning from 350 to 650$\,\degree\text{C}$. Circle, square, and diamond markers are added for clarity on top of the isotherms corresponding to pure As, Sb, and Sn in the  $\Sigma 3(1\bar{1}1)$ (solid lines), $\Sigma 3(1\bar{1}2)$ (dashed lines), and $\Sigma 9(2\bar{2}1)$ (dash-dotted lines), respectively.}
    \label{fig:mclena}
\end{figure}

The isotherms of As, Sb, and Sn as a function of temperature and composition are shown in Fig.~\ref{fig:mclena}. In these calculations, the concentrations of tramp elements As, Sb, and Sn were set to $0.05\,\textrm{wt.}\%$, while the alloying elements Mo, Ni and Cr were set to $2\,\mathrm{wt}.\%$. Since RTE is known to occur within a temperature range of 350$\degree\text{C}$ to 650$\degree\text{C}$, we limited the temperature range for plotting accordingly. The isotherms of the pure elements (As, Sb, and Sn) clearly exhibit a dependency on GB type. For the $\Sigma 3(1\bar{1}1)$ and $\Sigma 9(2\bar{2}1)$ GBs, similar GB occupancies are maintained across the entire temperature range, while GB enrichment in the $\Sigma 3(1\bar{1}2)$ is notably reduced\add{, leading to practically no impact of potential segregation in this type of GB on the mechanical properties.}. This observation can be attributed to reduced segregation trapping within this GB type (cf.~\ref{fig:single_segreg_mech_properties}), combined with a lower driving force for segregation at elevated temperatures. When a second element is introduced, the final GB occupancy is primarily determined by the differences in the individual elements' segregation tendencies to specific GB sites. Consequently, when the segregation energies of two components to the same GB site are closely matched, site competition becomes more pronounced. This can be exemplified by the case of As with Ni or As with Mo in the $\Sigma 3(1\bar{1}1)$ GB. Although As and Ni exhibit a similar ``up and down" pattern in their segregation profiles (Fig.\ref{fig:seg_profiles}a), $E_{seg}$ of Ni has $\approx 3\times$ smaller magnitude than that of As. On the contrary, Mo competes more strongly with As due to their similar segregation strengths to site 1, resulting in a greater reduction in GB enrichment when Mo is added compared to Ni. Conversely, in the $\Sigma 9(2\bar{2}1)$ GB, As and Ni show their second highest segregation tendency to site 4 (Fig.\ref{fig:seg_profiles}c), with As and Ni experiencing stronger site competition compared to As and Mo, leading to a more significant reduction in the GB enrichment of As with Ni compared to Mo. Overall, without delving into excessive detail on every elemental combination, it is worth noting that the site competition of Cr with As, Sb, and Sn is too weak to exert a significant impact on the GB enrichment of these tramp elements within this temperature range. This primarily stems from Cr being a weak segregant in the investigated GBs. Ni consistently tends to lower GB enrichment in all cases but to varying degrees, while Mo only exhibits a noticeable effect on As in the $\Sigma 3(1\bar{1}1)$ GB and on Sb in the $\Sigma 9(2\bar{2}1)$ GB. However, the most substantial site competitions are observed among the tramp elements themselves, reflecting their high overall segregation strengths across the segregation profile.

\begin{figure}[htp]
    \centering
    \includegraphics[width=8cm]{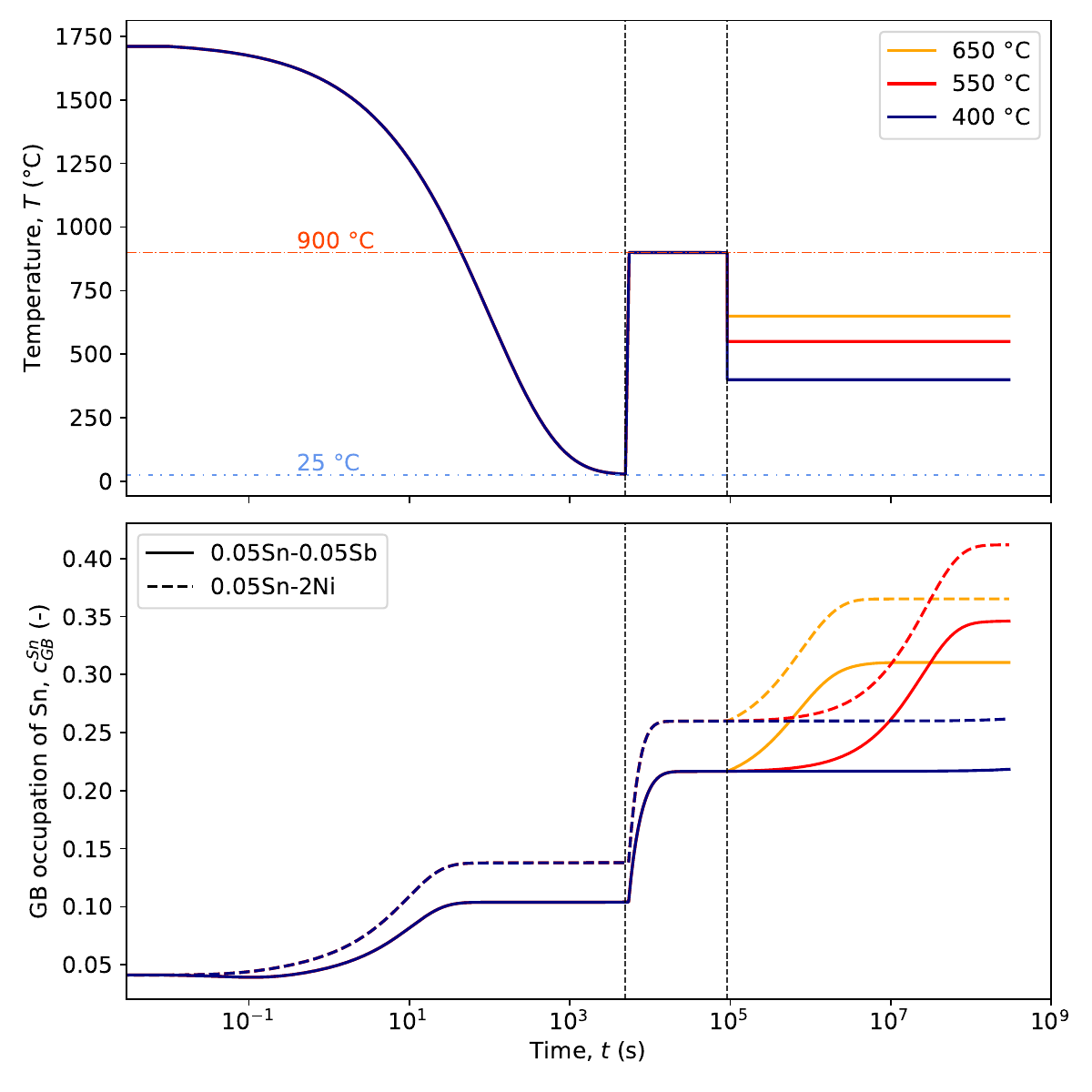}
    \caption{GB enrichment of Sn in the \add{$\Sigma 3(1\Bar{1}1)$ GB} for two different alloy compositions in \add{$\textrm{wt.}\%$} (bottom panel) and three different isothermal heating temperatures (top panel).}
    \label{fig:kinetic}
\end{figure}

The site competition effect becomes particularly evident in the kinetic simulation of the GB enrichment of Sn, as illustrated in Fig.~\ref{fig:kinetic}. Notably, the enrichment of Sn is more pronounced when coexisting with $2\,\text{wt.\%}$ Ni compared to its coexistence with $0.05\,\text{wt.\%}$ Sb. Furthermore, the results reveal that the final GB enrichment of Sn increases as the isothermal temperature decreases from 650$\,\degree\text{C}$ to 550$\,\degree\text{C}$. When observing the isotherms of Sn in conjunction with Sb or Ni, as depicted in Fig.~\ref{fig:mclena}, it becomes apparent that the GB enrichment obtained from the kinetic simulation converges towards the final values shown in the isotherms as time approaches infinity ($t\rightarrow\infty$). Consequently, it can be inferred that the GB enrichment in the kinetic simulation at a temperature of 400$\,\degree\text{C}$ would be even higher compared to the values obtained at 550$\,\degree\text{C}$ and 650$\,\degree\text{C}$, respectively. However, due to the significantly lower temperature, the kinetics are substantially slowed down, resulting in only marginal increases even after $300\cdot10^6\,$s. \add{For completeness, the enrichments of Sb and Ni are given in Supplementary Figure S3.}

\section{Discussion}
\subsection{Phosphorous vs. other tramp elements}
Among the diverse elements investigated in this study, it is unequivocally evident that Sn, Sb, and As stand out with the highest segregation energies across all GB types. Furthermore, these elements show a profoundly detrimental influence on GB cohesion and lead to intergranular fracture in all investigated GBs. Hence, based on their pronounced segregation tendencies, it is not coincidental that these elements are detected during measurements of fractured surfaces, such as by Auger electron spectroscopy~\cite{ohtani1976temper, cianelli1977temper}. Surprisingly, the effect of As on GB cohesion is on par with that of Sn and Sb in all investigated GB types\del{,}\add{.} \del{despite the prior belief that the impact of As is less harmful than that of Sb and Sn~\mbox{\cite{seah1975interface, seah1976segregation}}}. \add{This finding contradicts the pair bonding based theoretical predictions by Seah~\etal~\cite{seah1975interface, seah1976segregation} that As is less detrimental compared to Sb and Sn.}

Semi-empirical equations describing the dependence of the transition temperature shift ($\Delta T_T$) from ductile to brittle failure of alloying elements~\cite{seah1976segregation}, e.g., $\Delta T_T = 0.28\text{P}+0.38\text{Sb}+0.16\text{Sn}+0.048\text{As}$, (in weight percent) \add{may have} underestimated the influence of As. This discrepancy \del{is likely due to} \add{could be related to two factors: First, significant solute interactions with other structural defects such as dislocations, which we did not account for; and/or second,} the relatively low As contents in the investigated steels. However, with the potential increase in recycling rates, these semi-empirical descriptions may require adjustment in the future. 

Furthermore, a direct comparison of the impact of As, Sb, and Sn on GB cohesion with that of P, taken from Ref.~\cite{mai2023phosphorus}, in the investigated GB types revealed that their impact exceeds that of P in all cases. \add{Specifically, in the $\Sigma 3(1\Bar{1}1)$ GB, the detrimental effect increases by 0.39 J/m$^2$, 0.53 J/m$^2$ and 0.29 J/m$^2$, respectively. In contrast, within the $\Sigma 9(2\Bar{2}1)$ GB, the effect is more pronounced, reaching an increase of 0.58 J/m$^2$, 0.58 J/m$^2$ and 0.39 J/m$^2$ for As, Sb and Sn, respectively.}
\del{Particularly, in the two high-angle GBs, the detrimental impact is approximately two times greater in the $\Sigma 3(1\bar{1}1)$ and roughly sixteen times greater in the $\Sigma 9(2\bar{2}1)$ GB.} Although the impact difference varies depending on the GB type, a clear trend is indisputable, demonstrating that these elements contribute more significantly to RTE than P does. 

\subsection{Co-segregation and RTE}
As proposed in previous literature~\cite{guttmann1975equilibrium}, Mo reduces proneness to RTE through a scavenging effect resulting from attractive bulk interactions. However, our calculations indicate repulsive interactions between Mo and combinations with As, Sb, and Sn. Consequently, these interactions could potentially promote the segregation of the tramp elements by separating the elements from each other. Furthermore, only Ni attracts As, Sb, and Sn in bulk, with attractive interactions in the first nn-shell, albeit of small magnitude ($<0.09\,\text{eV}$), and repulsive interactions from the second nn-shell onward. Interestingly, the most significant repulsion occurs between the tramp elements themselves, thus promoting their separation to greater distances within the bulk.

Conversely, the interactions at the GBs significantly diverge from those in the bulk, ranging from $-0.32\,\text{eV}$ to $0.62\,\text{eV}$, exhibiting a notable dependence on the GB type. The most pronounced attractive interactions are observed between Ni--Sb and Ni--Sn pairs in the $\Sigma 3(1\bar{1}2)$ GB, followed by the Sb--Sb, Sb--Sn, and Sn--Sn interactions in the $\Sigma 3(1\bar{1}1)$ GB. Consequently, the co-segregation energies of these elements are reduced, signifying a higher segregation tendency. Assessing the cohesion change of Ni--Sb and Ni--Sn pairs in the $\Sigma 3(1\bar{1}2)$ GB (Fig.~\ref{fig:multi_coseg_mech_properties}b), we observe a notably adverse combined impact, although this combined effect is only marginally different from the individual impacts of Sb and Sn. The remaining interactions between Ni and the tramp elements are mostly slightly repulsive, with values ranging from $-0.10\,\text{eV}$ to $0.11\,\text{eV}$. Consequently, the co-segregation is not significantly affected. Similar to the previous scenario, the additional effect of Ni on cohesion is negligible in most cases, with the exception of the $\Sigma 9(2\bar{2}1)$. In this case, the detrimental effect of As, Sb, and Sn on cohesion is further increased by 25$\,\%$, 13$\,\%$, and 28$\,\%$, respectively. 
The attractive interactions observed between Sn/Sb combinations in the $\Sigma 3(1\bar{1}1)$ GB even result in doubling the negative impact on cohesion (Fig.~\ref{fig:multi_coseg_mech_properties}a). However, the majority of the remaining interactions between the tramp elements are repulsive, which diminishes the co-segregation tendency. In the $\Sigma 3(1\bar{1}1)$ and $\Sigma 9(2\bar{2}1)$ GBs, all elemental combinations continue to display a co-segregation tendency, with energies ranging from $-0.49$\,\text{eV}$ $ to $-0.97\,\text{eV}$. Contrastingly, only two combinations, specifically the segregation of Sn following prior segregation of Sb or Sn, sustain a segregation tendency in the $\Sigma 3(1\bar{1}2)$ GB.

The consequences of these co-segregations on cohesion are twofold; in most cases, the detrimental cohesion change is doubled, while Sn/Sb combinations in the $\Sigma 9(2\bar{2}1)$ GB exhibit a less cohesion-detrimental effect compared to the single element effect, although still leading to embrittling, ranging from $-0.35\,$J/m$^2$ to $\add{-}0.51\,$J/m$^2$.
Strong repulsive interactions are observed between Mo and As in all GB types, as well as between Mo and Sb and Mo and Sn in the $\Sigma 3(1\bar{1}2)$ and $\Sigma 9(2\bar{2}1)$ GBs. Based on our calculations, Mo, much like Cr, is found to be anti-segregating in the $\Sigma 3(1\bar{1}2)$ GB, and due to the repulsive interactions with As, Sb, or Sn, it remains anti-segregating also during the co-segregation scenario. Consequently, there is no substantial point in discussing the co-effect on the cohesion change of these elemental pairs. The cohesion change is influenced by the effect of pure As, Sb, or Sn, which are all strongly embrittling.

As discussed in section~\ref{cosegs_ints}, the order of segregation can play a pivotal role. Therefore, Mo is capable of reducing the segregation strength of the tramp elements in the $\Sigma 9(2\bar{2}1)$ GB if it segregates first but is completely hindered from segregating if As, Sn, or Sb segregates first. In the former case, where the tramp elements still exhibit pronounced segregation tendencies in the range of $-0.40\,\text{eV}$ to $-0.57\,\text{eV}$, the final cohesion change is governed by a scenario where Mo and one of the tramp elements are present at the GB. As indicated in Fig.~\ref{fig:multi_coseg_mech_properties}c, the impact of As is almost diminished, while the detrimental impact of Sb and Sn is reduced by approximately 30$\,\%$ and 50$\,\%$, respectively. In the latter case, where the segregation of Mo is prevented from segregating due to repulsive interactions, the cohesion change is controlled solely by the tramp elements. \add{We note that this conclusion is based on our highly concentrated DFT models. Therefore, the segregation of Mo will only be completely hindered if a high tramp element coverage is already present at the GBs. Given the fact that the amount of Mo is typically orders of magnitude higher than the amount of Sn, Sb, or As, we expect a mixed scenario, with regions occupied by Mo and regions occupied by tramp elements.}

In the $\Sigma 3(1\bar{1}1)$ GB, where small negligible attractive interactions exist between Mo and Sb, and Mo and Sn, all elements can enrich at the GB. Consequently, the negative effect of Sb and Sn on the cohesion is reduced by approximately 60$\,\%$ and 80$\,\%$, respectively. The same trends can be observed for the co-segregation of Cr. Nevertheless, Cr exhibits weaker repulsive forces on the incoming tramp elements, and its impact on the cohesion is thus negligible. Therefore, the cohesion change in Cr-tramp element combinations remains essentially unaffected. 
Based on our energetic and interaction evaluations,  there is no direct evidence that Ni and Cr promote the GB enrichment of Sn or Sb, which contrasts previous literature~\cite{ohtani1976temper, cianelli1977temper}. The only strong attractive interactions that could potentially promote GB enrichment are predicted between Ni--Sn and Ni--Sb in the bulk-like $\Sigma 3(1\bar{1}2)$ GB. However, it is essential to consider that in real alloys, the majority of GBs consist of general GBs, which are in our study represented by the $\Sigma 3(1\bar{1}1)$ and $\Sigma 9(2\bar{2}1)$ GBs exhibiting local non-bulk-like configurations.

Moreover, our thermodynamic simulations (Fig.~\ref{fig:mclena}), suggest that Sb exhibits no significant enrichment, and Sn only enriches to a small extent (from 0.025\,at.\% in bulk to approximately 6\,at.\% at the GB) at 350$\,\degree\text{C}$. Analyzing the other isotherms for the $\Sigma 3(1\bar{1}1)$ and $\Sigma 9(2\bar{2}1)$ GBs, it becomes evident that Ni tends to reduce the GB enrichment of As, Sb, and Sn due to site competition effects. These findings, in combination with our calculated, mostly repulsive interactions, lead to the conclusion that Ni does not enhance GB enrichment of tramp elements in real alloys. Instead, in most of the cases studied here, it tends to reduce the final GB coverage of the tramp elements while enriching to a lesser extent. Consequently, the final change in cohesion is governed by the combined effects of Ni and the tramp elements. However, this impact is negligible compared to the influence of pure As, Sb, and Sn, particularly in the $\Sigma 3(1\bar{1}1)$ GB, and thereby it contributes to cohesion reduction in the $\Sigma 9(2\bar{2}1)$ GB.

In contrast, due to the much smaller segregation tendency of Cr to $\Sigma 3(1\bar{1}1)$ and $\Sigma 9(2\bar{2}1)$ GBs, we observe no significant impact on the isotherms of the tramp elements. In this temperature range, the GB concentrations of As, Sb, and Sn can develop alongside Cr without restrictions. Even with Cr at the GBs, no substantial influence on the change in cohesion is evident. Therefore, the experimentally observed shifts in the transition temperature of Cr-containing alloys are most likely attributed to the formation of carbides, which can independently alter GB cohesion. Additionally, the absence of C, known to enhance GB cohesion and displace detrimental elements from the GB ~\cite{seah1975interface, seah1976segregation}, allows As, Sb, and Sn to enrich. This provides an explanation for the observed shift in the transition temperature in Cr-containing alloys. 

The situation for Mo is somewhat clearer. All calculated results show that Mo is a potent cohesion enhancer when segregated at the GB and not because of scavenging effects. Looking at the isotherms of As, Sb, and Sn in the non-bulk-like $\Sigma 3(1\bar{1}1)$ and $\Sigma 9(2\bar{2}1)$ GBs one can see that Mo exhibits enough site competition with As in the $\Sigma 3(1\bar{1}1)$ and with Sb in the $\Sigma 9(2\bar{2}1)$ GB to reduce the enrichment. The strong repulsive interactions which Mo exhibit on the elements when first segregated, especially in the $\Sigma 9(2\bar{2}1)$ GB, could lead to further reduction of the GB enrichment of As, Sb, and Sn in the critical temperature region for RTE. Hence, increasing Mo content in RTE-vulnerable steels definitively mitigates the embrittlement. 

Finally, we demonstrate that companion elements exhibit the most substantial influence on their respective isotherms when interacting with each other. This phenomenon is primarily attributed to their robust competition for sites, driven by their equally strong segregation tendencies. This is evident in the kinetic simulation within the $\Sigma 3(1\bar{1}1)$ GB, where, despite competing with $2\,\text{wt.\%}$ Ni, the GB enrichment of Sn is approximately 6$\,\%$ higher at 650$\,\degree\text{C}$ and 550$\,\degree\text{C}$ compared to the presence of $0.05\,\text{wt.\%}$ Sb. However, it is worth noting that Sb will enrich alongside Sn at the GB, thereby affecting cohesion in a combined manner.

In summary, our calculations suggest that the combinations of tramp elements have a considerably detrimental impact on cohesion in all examined cases and, in most instances, lead to more significant GB embrittlement than the individual species alone.

\section{Conclusions}
The individual and combined effects of As, Sb, Sn, Ni, Cr, and Mo segregation, as well as their relationship to reversible temper embrittlement (RTE), have been systematically investigated in three high-angle bcc Fe tilt GBs, specifically the $\Sigma 3 (1\bar{1}1)[110]$, $\Sigma 3(1\bar{1}2)[110]$, and $\Sigma 9(2\bar{2}1)[110]$ GBs. In particular, we examined the effects of segregation and co-segregation on GB cohesion, discussed the interactions between these elements at the GBs, and compared them to their interactions in the bulk. To gain a deeper understanding of the GB enrichment of As, Sb, and Sn within the critical temperature range for RTE, which occurs between 350$\,\degree\text{C}$ and 650$\,\degree\text{C}$, we employed a series of consecutive thermodynamic and kinetic simulations, considering the multi-component and multi-site nature of these systems. The primary findings from our investigations can be summarized as follows:  

\begin{enumerate}
    \item Sn, Sb, and As exhibit strong segregation tendencies and substantially reduce GB cohesion, promoting intergranular fracture.
    \item Their detrimental effect on the GB cohesion is ordered as Sb$>$As$>$Sn. In contrast to the original works of M.P.~Seah~\cite{seah1975interface, seah1976segregation} on the impact of solute segregation on GB strength, our results suggest that As promotes GB decohesion slightly more than Sn does, although the differences between Sb, As, and Sn are rather small.
    \item  The solute-solute interactions in bulk are primarily repulsive, suggesting that Mo does not impede RTE by scavenging As, Sb, and Sn. Instead, Mo's beneficial influence stems from its capacity to enhance cohesion when segregated to GBs. Moreover, it exhibits repulsive interactions with tramp elements, and depending on the GB type and tramp element, it competes effectively for sites, thus reducing the overall tramp element enrichment. Consequently, by increasing the Mo content, the impact of temper embrittlement can be significantly mitigated.
    \item In general, Ni and Cr do not contribute to increased GB enrichment of As, Sb, and Sn, as their interactions are predominantly weak to moderately repulsive. However, attractive interactions exist between Ni and Sb, as well as Ni and Sn, in the bulk-like $\Sigma 3(1\bar{1}2)[110]$ GB. Nonetheless, co-segregation of Cr with tramp elements does not significantly affect GB cohesion when compared to the single-element segregation of tramp elements. Therefore, the susceptibility of Cr-containing low-alloy steels to RTE likely arises from the formation of Cr-carbides, which can detrimentally impact GB cohesion, and the removal of C provides space for cohesion-deteriorating elements to segregate. Similar to Cr, Ni has a negligible effect on cohesion when co-segregated. Only in the $\Sigma 9(2\bar{2}1)[110]$ GB do we consistently observe that cohesion worsens with the co-segregation of Ni. 
    \item The interactions between the tramp elements are mostly repulsive, although strong attractive interactions occur between Sn and Sb combinations in the $\Sigma 3 (1\bar{1}1)[110]$ GB. Despite this, the co-segregation tendency of these elements still remains remarkable in the general/high-angle type GBs, ranging from $-0.47\,$eV to $-0.97\,$eV. This should be taken into account above all by the planned future increase in the use of scrap in steel production. In addition, it could be shown that the influence of co-segregation reduces cohesion in most cases twice as much as compared to single-species segregation. 
\end{enumerate}

\section*{Acknowledgments}
The financial support by the Austrian Federal Ministry for Digital and Economic Affairs, the National Foundation for Research, Technology and Development and the Christian Doppler Research Association is gratefully acknowledged.
\newpage
\appendix

\newpage
 \bibliographystyle{elsarticle-num} 
 \bibliography{cas-refs}





% \end{thebibliography}
\end{document}